\documentclass[aps, prd, twocolumn, amsmath, floats,floatfix, superscriptaddress, nofootinbib]{revtex4}
\usepackage{graphicx}
\usepackage{bm}
\usepackage{diagbox}
\usepackage{hyperref}
\usepackage[utf8]{inputenc}
\usepackage{subfigure,booktabs}
\usepackage{pifont}
\makeatletter\def\Hy@Warning#1{}\makeatother
\hypersetup{
  colorlinks=true,        
  linkcolor=blue,         
  citecolor=cyan,         
  urlcolor=cyan            
}
\usepackage{amsmath,amsfonts,amssymb}
\usepackage{times}
\usepackage[T1]{fontenc}
\usepackage{float}
\usepackage[normalem]{ulem}
\usepackage{multirow}
\usepackage{soul}
\usepackage[justification=raggedright,singlelinecheck=false]{caption}

\newcommand{\be}{\begin{equation}}
\newcommand{\ee}{\end{equation}}
\newcommand{\bea}{\begin{eqnarray}}
\newcommand{\eea}{\end{eqnarray}}

\makeatletter
\newcommand*{\rom}[1]{\expandafter\@slowromancap\romannumeral #1@}
\makeatother

\makeatletter
\newcommand{\thickhline}{%
    \noalign {\ifnum 0=`}\fi \hrule height 1pt
    \futurelet \reserved@a \@xhline
}
\newcolumntype{"}{@{\hskip\tabcolsep\vrule width 1pt\hskip\tabcolsep}}
\makeatother
\begin{document}
\title{
Neural Posterior Estimation of Neutron Star Equations of State

}

\author{Valéria Carvalho}
\email{val.mar.dinis@uc.pt}
\affiliation{CFisUC, 
	Department of Physics, University of Coimbra, P-3004 - 516  Coimbra, Portugal}

\author{Márcio Ferreira}
\email{marcio.ferreira@uc.pt}
\affiliation{CFisUC, 
	Department of Physics, University of Coimbra, P-3004 - 516  Coimbra, Portugal}

\author{Micha{\l} Bejger}
\email{bejger@camk.edu.pl}
\affiliation{INFN Sezione di Ferrara, Via Saragat 1, 44122 Ferrara, Italy}
\affiliation{Nicolaus Copernicus Astronomical Center, Polish Academy of Sciences, Bartycka 18, 00-716, Warsaw, Poland}

\author{Constança Providência}
\email{cp@uc.pt}
\affiliation{CFisUC, 
	Department of Physics, University of Coimbra, P-3004 - 516  Coimbra, Portugal}
\date{\today}

\begin{abstract} 
We present a simulation-based inference framework to constrain the neutron star (NS) equation of state (EoS) from astrophysical observations of masses, radii and tidal deformabilities, using neural posterior estimation with conditional normalizing flows. To ensure that the model conforms with reality, physics-informed constraints are embedded directly into the training loss. This enables efficient, likelihood-free inference of full posterior distributions for key thermodynamic quantities—including pressure, squared speed of sound, and the trace anomaly—conditioned on observational data.
Our models are trained on synthetic datasets generated from two agnostic EoS priors: polytropic parameterizations and Gaussian process reconstructions. These datasets span various scenarios, including the presence or absence of tidal deformability information and observational noise. Across all settings, the method produces accurate and well-calibrated posteriors, with uncertainties reduced when tidal deformability constraints are included.
Furthermore, we find that the behavior of normalized predictive dispersions is strongly correlated with the maximum central density inside NSs, suggesting that the model can indirectly infer this physically meaningful quantity. The approach generalizes well across EoS families and accurately reconstructs derivative quantities such as the polytropic index, demonstrating its robustness and potential for probing dense matter in NS cores.

\end{abstract}

\maketitle

\section{Introduction}

Understanding the equation of state (EoS) of neutron star (NS) matter remains one of the central challenges in modern nuclear physics and astrophysics. The EoS governs the relationship between pressure and energy density at supranuclear densities, directly influencing a NS's internal structure and observable properties—most notably its mass, radius, and tidal deformability \cite{glendenning2012compact}. However, since these extreme conditions lie far beyond the reach of terrestrial experiments, the EoS must be constrained indirectly through a combination of astrophysical observations and nuclear-theory models \cite{chatziioannou2024neutron}.

Recent years have seen significant advances in observational capabilities. The detection of gravitational waves (GWs) from binary NS inspirals, particularly GW170817 \cite{Abbott:2018wiz} and GW190425 \cite{abbott2020gw190425}, has imposed strong constraints on tidal deformability, substantially reducing prior uncertainty in the mass–radius relation \cite{annala2018gravitational}. In parallel, X-ray pulse profile modeling from the Neutron star Interior Composition Explorer (NICER) mission has provided high-precision radius measurements for pulsars such as PSR J0030+0451 \cite{Riley_2019,Miller19} and J0740+6620 \cite{Fonseca:2021wxt}, finding radii near 12 km with uncertainties of ±0.5 km \cite{rutherford2024constraining}. These multimessenger measurements, when combined with low-density constraints from nuclear effective field theories like $\chi$EFT, have significantly narrowed the pressure bounds at supranuclear densities.
Recent NICER observations include PSR J0437–4715 \cite{choudhury2024nicer} and PSR J1231-1411 \cite{salmi2024nicer}.

Despite these breakthroughs, the high-density regime, which is greater than approximately twice nuclear saturation density, remains poorly constrained. The internal composition and stiffness of matter in this region remain uncertain, leaving open questions about the presence of phase transitions and exotic degrees of freedom, and the true nature of cold dense matter. Bridging this gap requires an integrated approach combining \textit{ab initio} calculations, agnostic modeling strategies, and Bayesian inference applied to multimodal observational data.

Traditional Bayesian frameworks, employing Markov Chain Monte Carlo (MCMC) methods, have been widely used to infer the EoS by sampling from the posterior distribution conditioned on observational constraints. However, these methods face practical limitations in high-dimensional or computationally expensive settings: slow convergence, difficulty in handling complex likelihoods, and limited scalability. To address these challenges, recent work has turned to machine-learning–based inference methods that bypass the need for an explicit likelihood function.

In particular, Neural Posterior Estimation (NPE) offers a compelling alternative. As a form of likelihood-free or simulation-based inference (SBI), NPE learns a direct mapping from observations to posterior distributions using a neural density estimator. This allows it to efficiently approximate complex, high-dimensional posteriors without requiring likelihood evaluations. NPE has already demonstrated success in applications such as GW parameter estimation \cite{dax2021real, Dax:2024mcn}, and it is well-suited to the NS EoS problem.
Within this framework, conditional normalizing flows (CNFs) have emerged as a powerful architecture for modeling flexible, invertible transformations between simple base distributions and complex target posteriors. Prior work has applied normalizing flows (NFs) to NS physics \cite{brandes2024neural,wouters2025leveraging,carvalho2024detecting}, although with different goals or inference targets.

In this work, we use NPE with CNFs to infer the EoS from synthetic observations, aiming to evaluate the model’s ability to recover the pressure $p(n)$, squared speed of sound $c^2_s(n) = \partial p(n)/\partial \epsilon(n)$, and trace anomaly \cite{fujimoto2022trace}, $\Delta(n) = 1/3 - p(n)/\epsilon(n)$, as functions of baryon density $n$, with $\epsilon(n)$ denoting the mass-energy density. To generate training data, we use two agnostic EoS priors: one based on polytropic parameterizations (PT), and the other on Gaussian process (GP) models \cite{2006gpml.book.....R}. Each prior is used to produce four datasets with different levels of observational realism: with or without tidal deformability constraints, and with or without injected noise. These variations allow us to study the impact of both measurement precision and available observables on the inference task. We also incorporate physics-based constraints into the CNF loss function, as introduced in \cite{DiClemente:2025pbl}, to enforce physically consistent behavior—such as the monotonic increase of pressure with density—directly during training.

Our results highlight the flexibility and robustness of the NPE-CNF framework. We show that including tidal deformability information significantly improves pressure estimates, while the squared speed of sound remains more sensitive to noise and harder to constrain. The model can also be used to reconstruct other quantities derived directly from the EoS, such as the polytropic index. An interesting result is that the increase in uncertainty associated with these properties at higher densities is correlated with the maximum central density inside a NS. This is a natural result, as the model does not receive information above the maximum central density from NS observations. On the other hand, it is a result that brings some information about the maximum central densities achieved inside NS.
Finally, we demonstrate that the trained models generalize well when tested on unseen EoS realizations, including a dataset constructed with phenomenological models not used during training. 
This work illustrates the potential of modern deep generative modeling to quantify uncertainty and extract physics from multisource NS data, laying the foundation for future applications to real observations and more expressive EoS parameterizations.
Despite significant progress from multimessenger observations—such as NICER's X-ray pulse profile modeling and gravitational-wave detections like GW170817—current data remain insufficient to fully constrain the high-density behavior and internal composition of NS matter. 
As shown in \cite{Huang:2023grj} and \cite{huxford2024accuracy}, present-day measurements primarily constrain the EoS at intermediate densities, largely reflecting the behavior of symmetric nuclear matter. However, they lack the discriminatory power needed to differentiate between competing high-density models, such as those involving purely nucleonic matter versus hyperonic or exotic compositions.
Additionally, recent efforts to detect potential first-order phase transitions using NICER data yield only weak Bayesian evidence \cite{huang2025constraining}, further emphasizing the limited sensitivity of current observations to the extreme-density regime.
Together, these findings suggest that resolving the EoS at densities beyond $\sim 2 n_0$ will require the greater precision and coverage expected from future missions such as STROBE-X \cite{ray2019strobe} and eXTP \cite{watts2019dense,zhang2019enhanced,li2025dense}.
In light of this, the present work does not rely on current observational data, but instead aims to simulate the precision of future measurements by generating synthetic datasets with reduced noise. 

The structure of this paper is as follows: Sec.~\ref{dataset} provides an overview of the datasets used, Sec.~\ref{NPE} introduces the NPE framework, CNFs, and implementation details,  Sec.~\ref{results} presents and discusses the results, and Sec.~\ref{conclusion} concludes the study.

\section{Dataset}
\label{dataset}

To train and evaluate the CNF inference framework developed in this work, we construct datasets derived from two distinct families of NS EoS: one parametric and one nonparametric. These base families are adapted to suit the specific requirements of our problem, allowing us to assess the model’s robustness across fundamentally different representations of dense matter.

\begin{itemize}
    \item \textbf{Polytropic EoSs (PT)} are generated following the piecewise-polytropic prescription in \cite{ferreira2025conditional}, which models the EoS as a sequence of connected polytropic segments. This approach offers a flexible and interpretable representation and is widely used in Bayesian inference of NS properties. 
    
    \item \textbf{Gaussian Process–based EoSs (GP)} are based on the nonparametric Bayesian framework developed in \cite{annala2023strongly, komoltsev_2023_10101447}. These models incorporate theoretical uncertainties and observational constraints without assuming a fixed functional form for the EoS.
\end{itemize}

The PT dataset from \cite{ferreira2025conditional} uses five-segment piecewise polytropes, allowing broad exploration of phase transitions and stiffness variations. Each EoS supports at least a $2\,M_\odot$ NS and satisfies causality (speed of sound in dense matter not larger than the speed of light in vacuum). Mass–radius relations are computed by solving the Tolman-Oppenheimer-Volkov (TOV) equations \cite{1939PhRv...55..364T,1939PhRv...55..374O}, and synthetic observational datasets are generated by sampling mass-radius pairs and perturbing them with Gaussian noise.

In contrast, the GP dataset in \cite{annala2023strongly} is constructed using GP regression to interpolate between well-constrained theoretical regimes: Chiral Effective Field Theory at low densities and perturbative QCD at high densities. The resulting EoS prior enforces causality, thermodynamic consistency, and conservative high-density behavior. This prior is then conditioned using Bayesian inference with current knowledge of highest NS masses from pulsars' observations,  \cite{antoniadis2013massive,cromartie2020relativistic}, tidal deformability measurement from the GW170817 event \cite{Abbott:2018wiz}, and simultaneous mass-radius measurements, by X-ray observations such as NICER \cite{shaw2018radius,steiner2018constraining,Fonseca:2021wxt}. 

The GP-based posterior ensemble reflects a broad yet physically motivated range of EoSs, including models with phase transitions and deconfined quark matter. Because it does not rely on a parametric form, the GP method provides a powerful test of the model’s ability to generalize across complex, model-agnostic EoS behavior.

This dual approach—spanning structured, interpretable PT models and flexible, constraint-driven GP ensembles—allows us to rigorously evaluate the model's performance and generalizability across the theoretical landscape of dense NS matter.

\subsection{Generation of Mock Datasets}\label{sec:ge_MD}

The size of the dataset for training, validation, and testing is as follows: 
\begin{itemize}
    \item \textbf{PT dataset}: 36,391 training+validation samples and 4,044 test samples.
    \item \textbf{GP dataset}: 70,176 training+validation samples and 7,798 test samples.
\end{itemize}
Both datasets follow a 90\%/10\% train–test split. Within the training and validation portion, we further allocate 90\% for training and 10\% for validation, resulting in final proportions of 81\% training, 9\% validation, and 10\% testing.
We acknowledge the discrepancy in dataset sizes and account for this during model evaluation. However, since our objective is not to directly compare the performance between PT- and GP-trained models, we do not apply dataset balancing or scale-dependent adjustments.\\

Due to differences in structure and origin, both datasets require preprocessing before training. In particular, for the GP dataset, we apply the following physical filtering criteria: {\it i)} the maximum mass must exceed $2\,M_\odot$; {\it ii)} the tidal deformability profile must exhibit a positive slope. We do not use likelihood-based filtering; only these physically motivated constraints are enforced.\\

The model inputs are vectors of pressure values evaluated at 20 fixed and equally spaced baryon density points:

\begin{equation}
    \boldsymbol{p} = [p(n_1), p(n_2), \cdots, p(n_{20})]\;,
\end{equation}
with \(n_1 = 0.13\,\text{fm}^{-3}\) and \(n_{20} = 1.28\,\text{fm}^{-3}\), where $n_{k+1}=n_k+\delta n$ with $\delta n =0.0605$, applied to both PT and GP datasets. For training, all pressure values are transformed to logarithmic scale to normalize their magnitudes and improve numerical stability during learning.

In addition to the pressure, we also analyze two key derivative quantities of the EoS that provide insight into the microscopic properties of dense matter: the squared speed of sound ($c_{s}^{2}$) and the trace anomaly ($\Delta$), though this analysis is performed only for the GP dataset.

The squared speed of sound, $c^{2}_{s}(n) = \partial p(n)/\partial\epsilon(n)$, where \( \epsilon(n) \) is the energy density, measures the stiffness of the EoS. Its value is crucial for determining the stability and causal structure of neutron star matter, with $c_s^2 \leq 1$ (in units of the speed of light) required by causality.

The trace anomaly, $\Delta(n) = \frac{1}{3} - p(n)/\epsilon(n)$, quantifies the deviation from conformal symmetry, a fundamental property of noninteracting massless particles. In the context of neutron star matter, a nonzero value signals the emergence of mass scales and strong interactions at high density, providing crucial insight into the microscopic degrees of freedom.

These quantities are represented on the fixed density grid as vectors:
\begin{align}
    \boldsymbol{c_s^2} &= [c_s^2(n_1), c_s^2(n_2), \cdots, c_s^2(n_{20})]\;, \\
\boldsymbol{\Delta} &= [\Delta(n_1), \Delta(n_2), \cdots, \Delta(n_{20})] \;.
\end{align}

We adopt a grid of 20 fixed baryonic density values spanning the relevant physical regime. This choice provides sufficient resolution to capture key EoS features while maintaining computational efficiency during training and inference.

\begin{figure}[!hbt]
    \centering
    \includegraphics[width=0.8\linewidth]{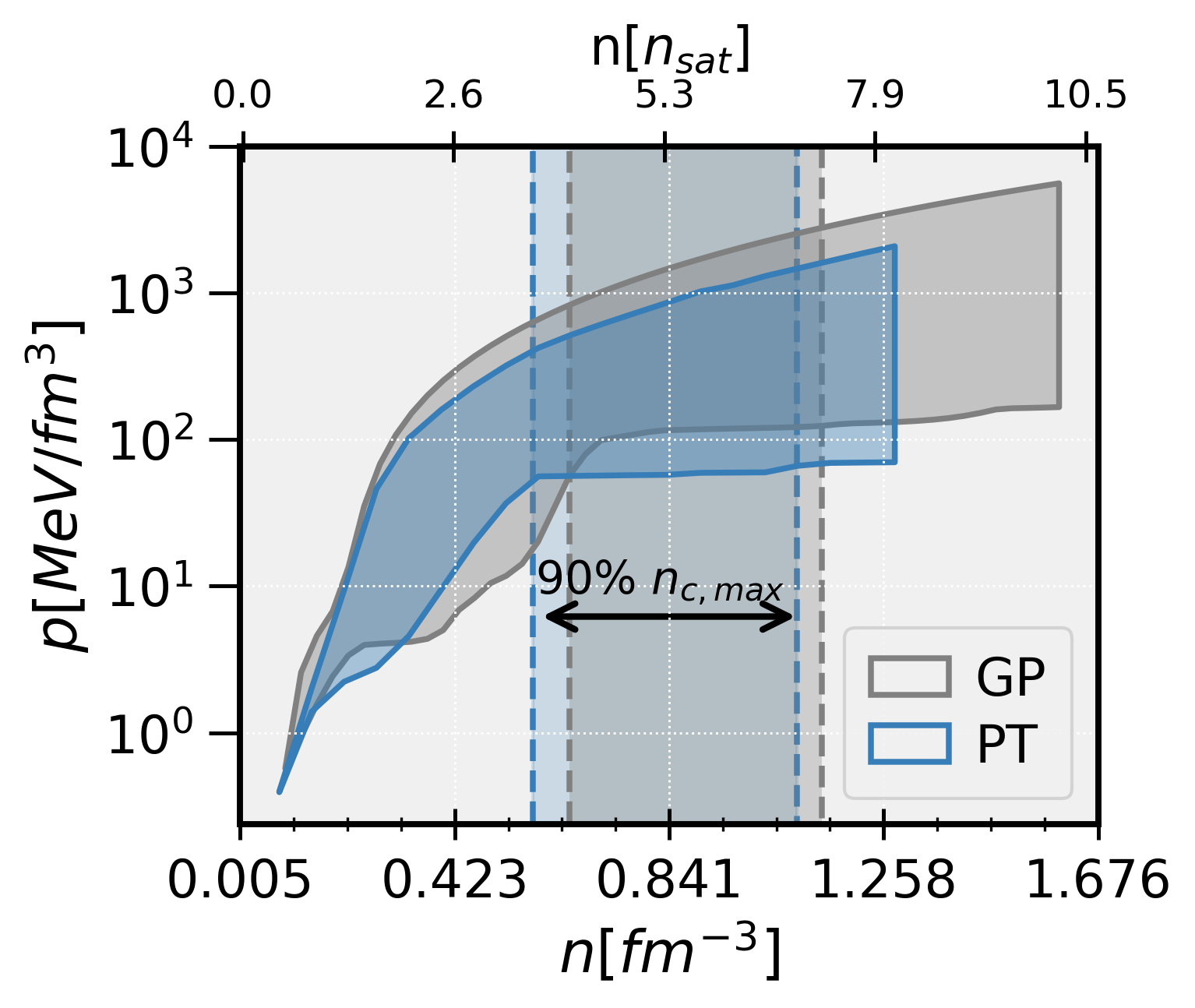}\\
    \includegraphics[width=0.8\linewidth]{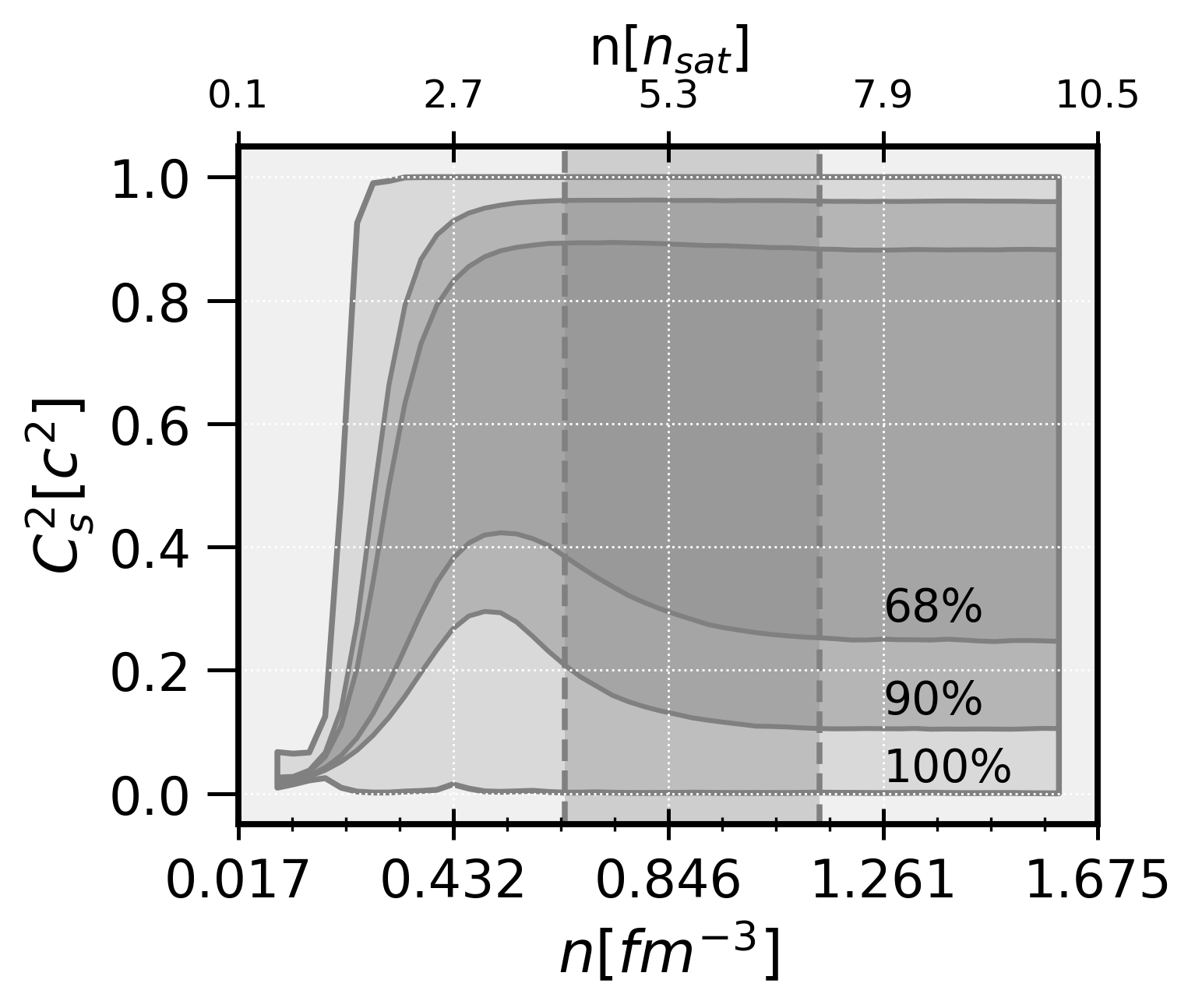}\\
        \includegraphics[width=0.8\linewidth]{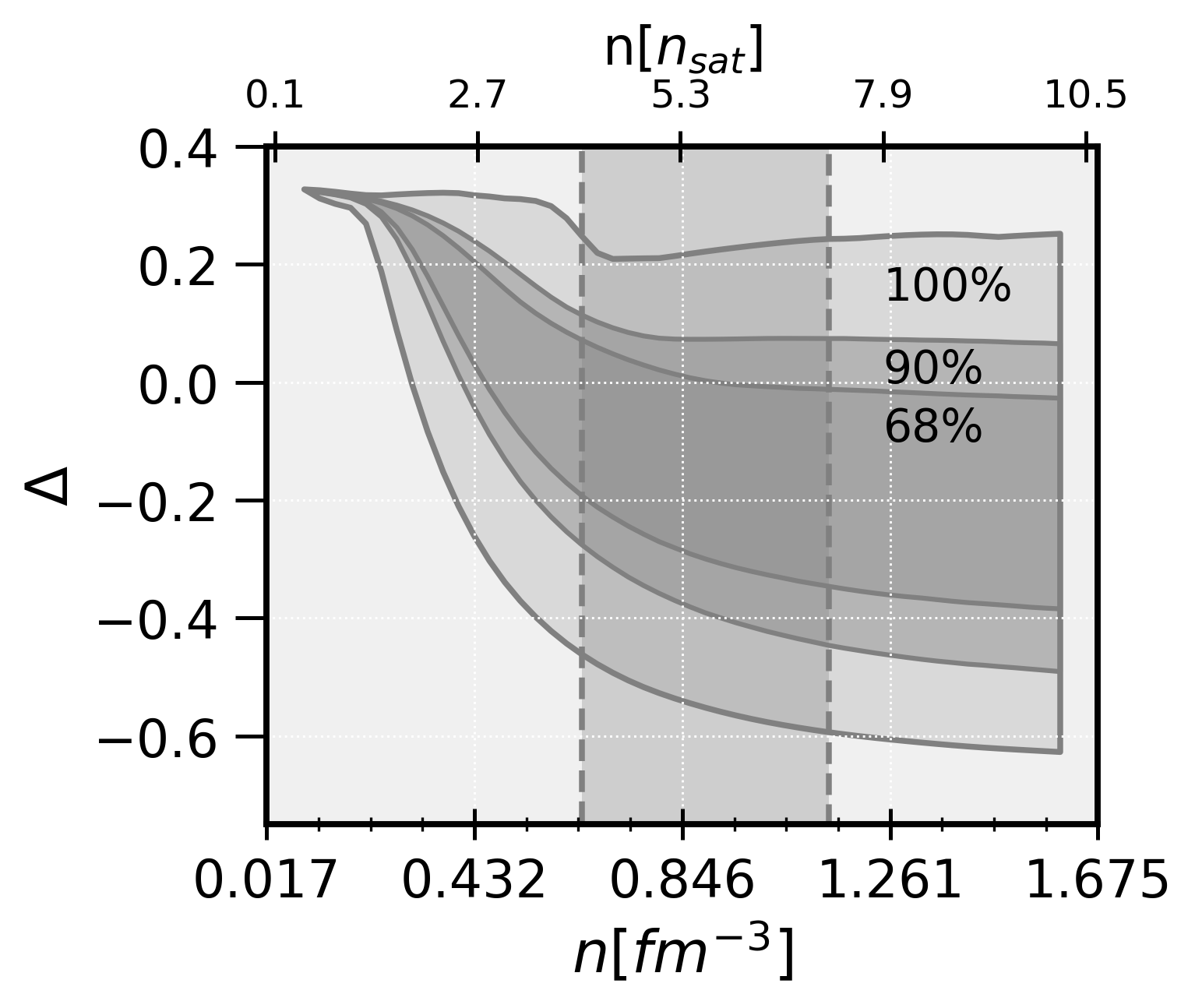}
    \caption{
    \textbf{Top:} Pressure versus baryonic density. Gray curves correspond to the GP dataset; blue curves represent the PT dataset.  
    \textbf{Middle:} Squared speed of sound versus baryonic density, shown only for the GP dataset.  
    \textbf{Bottom:} Trace anomaly versus baryonic density, shown only for the GP dataset.  
    In all plots, the vertical shaded region marks the 90\% quantile range of the maximum central density across the dataset.  }
    \label{fig:p_vs}
\end{figure}

Figure~\ref{fig:p_vs} illustrates the pressure density, speed of sound, and trace anomaly versus baryonic density relations for both datasets. The gray curves correspond to the GP-generated EoSs, while the blue curves represent the PT-generated EoSs. The vertical band represents the 90\% quantile range of maximum central densities.

\begin{figure}[!hbt]
    \centering
    \includegraphics[width=0.9\linewidth]{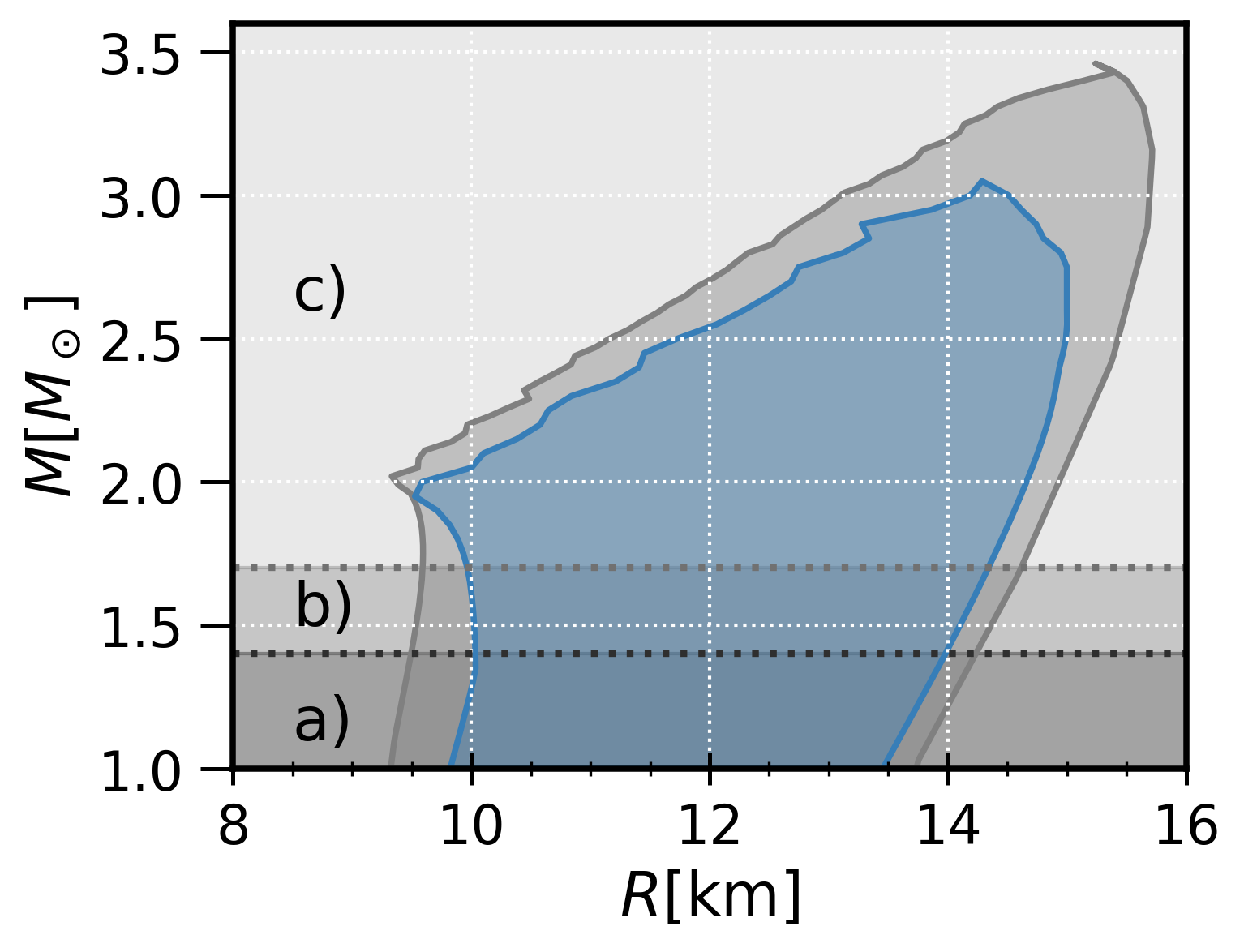}
    \includegraphics[width=0.9\linewidth]{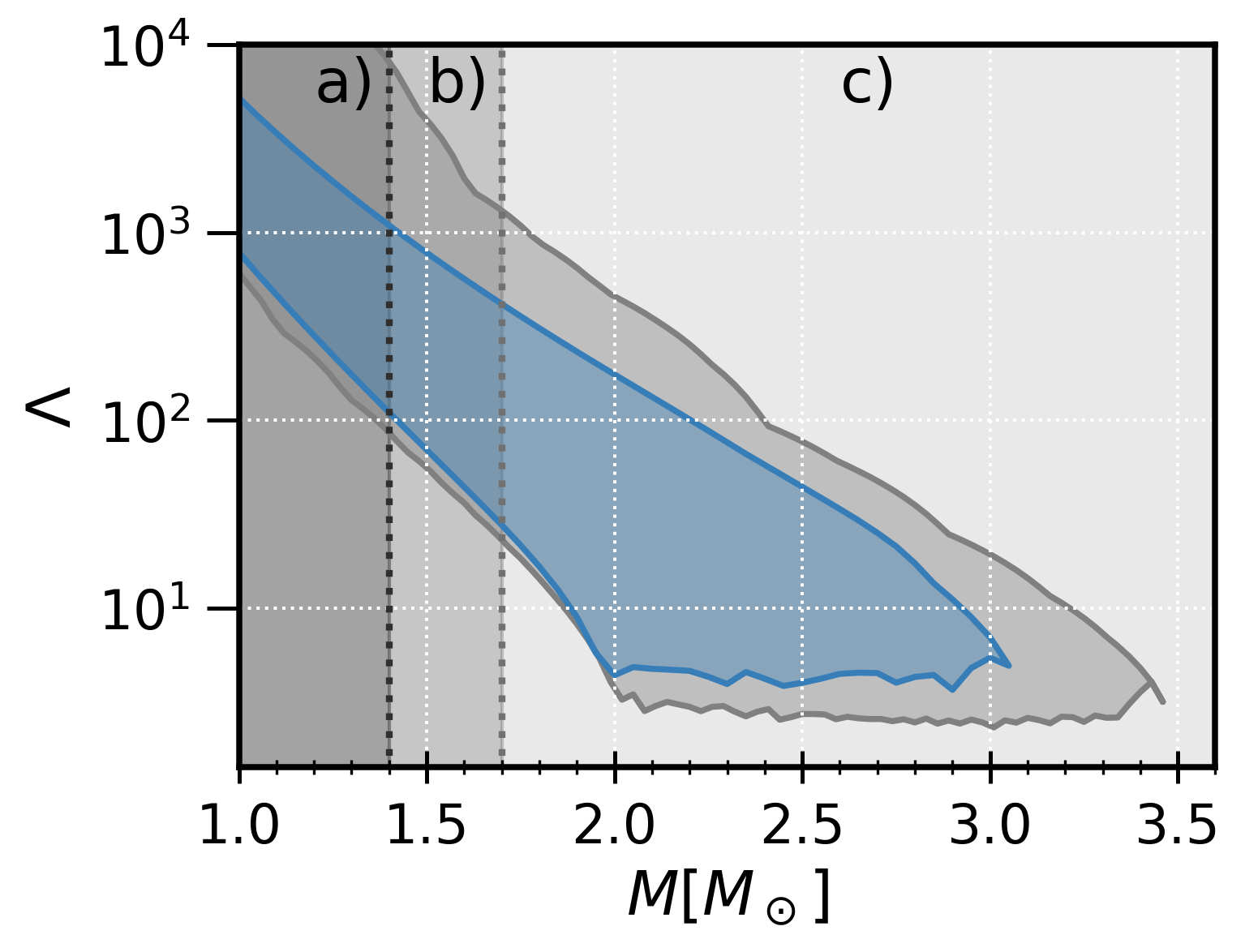}
\caption{ Range of mass-radius and mass-tidal deformability relations. The gray and blue regions show the band of the full extent of all relations generated from the GP and PT ensembles, respectively.
\textbf{Top:} Mass–radius relations.
\textbf{Bottom:} Mass–tidal deformability relations.
Gray curves correspond to the GP dataset; blue curves represent the PT dataset.
The shaded gray horizontal (top) and vertical (bottom) bands labeled a) from 1 to 1.4 $M_\odot$, b) from 1.4 to 1.7 $M_\odot$, and c) from 1.7 to $M_{\text{max}}(\text{EoS})$ $M_\odot$ denote the three mass ranges used in the uniform sampling procedure for mock data generation described in Sec. \ref{sec:ge_MD}}.
    \label{fig:M_r}
\end{figure}

To introduce observational noise and ensure robustness, we generated synthetic $M(R)$ observational data by first sampling 15 mass values from a uniform distribution. These samples were divided into three distinct regions with a ratio of [5,5,5]:

\begin{align}
    M^a & \in \mathcal{U}[1.0, 1.4]\,M_\odot \;,\\
    M^b &\in \mathcal{U}[1.4, 1.7]\,M_\odot \;, \\
    M^c  & \in \mathcal{U}[1.7, M_{\text{max}}(\text{EoS})]\,M_\odot \;, 
\end{align}
where $M_{\text{max}}(\text{EoS})$ represents the maximum mass supported by a given EoS, as determined by integrating the TOV equations and illustrated in Fig.~\ref{fig:M_r}. This three-region sampling strategy ensures that the model is exposed to a wide range of physically relevant masses, promoting better generalization across both well-populated and less-observed regimes. It is particularly useful for covering EoSs with softer high-density behavior—such as those including hyperons—which tend to support lower maximum masses.

These intervals are informed by astrophysical observations: most NSs are observed with masses between 1.4 and 1.7 $M_\odot$, with fewer objects below 1.4 or above 1.7 $M_\odot$. Although finer subdivisions were initially considered, they were found to decrease training efficiency due to excessive fragmentation.

To maintain consistency, the mass and corresponding radius values within each band are ordered from lowest to highest mass. This also enables a flexible strategy in which the network can be trained using only a subset of the $M(R)$ curve by zero-padding unused entries, which can improve focus on specific mass ranges. The undefined upper bound $M_{\text{max}}$ further supports the flexibility to apply additional physical constraints in future work.

During training, the mass vector takes the form

\begin{equation}
    \mathbf{M}^{0} = [M^{a}_1, \ldots, M^{a}_{n_o}, M^{b}_1, \ldots, M^{b}_{n_o}, M^{c}_1, \ldots, M^{c}_{n_o}]\;,
\end{equation}
where $n_o = 5$ denotes the number of samples per mass range, ensuring a uniform 1:1:1 ratio across all bands. The corresponding radius vector is then given by

\begin{align}
\mathbf{R}^{0} = &[\, R(M^a_1), \ldots, R(M^a_{n_o}), R(M^b_1), \ldots, R(M^b_{n_o}), \nonumber \\
                  &\quad R(M^c_1), \ldots, R(M^c_{n_o})\, ].
\end{align}

When incorporating tidal deformability, the vector representation is expanded: another mass vector is resampled with the same approach as before,
\begin{equation}
    \mathbf{M}^{*} = [M^{*a}_1, \ldots, M^{*a}_{n_o}, M^{*b}_1, \ldots, M^{*b}_{n_o}, M^{*c}_1, \ldots, M^{*c}_{n_o}] \; ,
\end{equation}
and the corresponding tidal deformability vector is given by
\begin{align}
\mathbf{\Lambda}^{0}  = [\Lambda(M^{*a}_1), \ldots, \Lambda(M^{*a}_{n_o}), \Lambda(M^{*b}_1), \ldots, \Lambda(M^{*b}_{n_o}), \nonumber \\
                \Lambda(M^{*c}_1), \ldots, \Lambda(M^{*c}_{n_o})].
\end{align}

In this work, we assume a common mass distribution for both the X-ray (M–R) and GW (M-$\Lambda$ ) observations. While the true mass distributions for isolated pulsars and merging binary NSs may differ, current observational data do not provide strong constraints to justify distinct priors. Adopting a unified mass distribution reflects the assumption that, in the absence of clear evidence to the contrary, these systems originate from the same underlying population. This choice simplifies the inference setup and allows for a more controlled comparison of how different observables constrain the EoS.

\begin{figure*}[!hbt]
    \centering
    \includegraphics[width=1\linewidth]{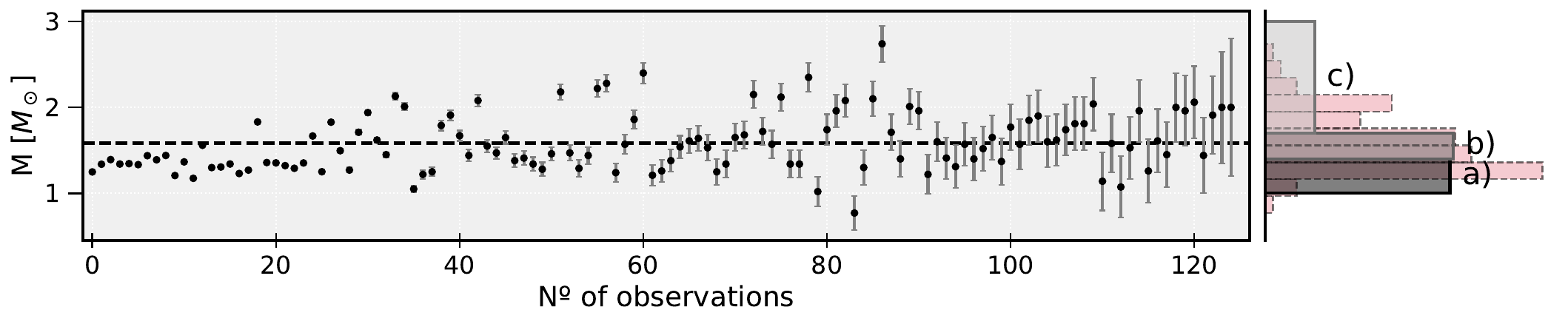}
    \caption{ Compilation of neutron star mass measurements from \cite{rocha2023mass}, supplemented with additional entries from J0030+0451 \cite{Riley_2019,Miller19}, J1231-1411 \cite{salmi2024nicer}, and J1731-347 \cite{doroshenko2022strangely}. Black points with error bars denote the reported central values and uncertainties of individual measurements. The dashed horizontal line marks the mean of all central mass values, $1.58,M_\odot$. The pink histogram on the right shows the distribution of these central values. The three shaded histogram bars, labeled a), 
    b), and c), indicate the mass intervals adopted for uniform sampling in this work (see Sec.~\ref{sec:gm_test}).}
    \label{fig:Mass_distr}
\end{figure*}

To emulate realistic observational uncertainties, Gaussian noise is applied to the ``ground truth'' (TOV-computed) values. Each observed quantity is sampled from a normal distribution centered at its true value with a variance drawn from a uniform distribution over a prescribed range.
For mass and radius, we have
\begin{align}
    \mathbf{M}  \sim \mathcal{N} (\mathbf{M}^0 ,\mathbf{\sigma_M^2})\;,  \qquad & \mathbf{R} \sim \mathcal{N} (\mathbf{R}^{0} ,\mathbf{\sigma_R^2}) \;,\\
    \mathbf{\sigma_{M}} \sim \mathcal{U}[0,\sigma_M]\;, \qquad   & \mathbf{\sigma_R} \sim \mathcal{U}[0,\sigma_R]
    \;,
\end{align}
and for tidal deformability
\begin{align}
\mathbf{\Lambda}\sim \mathcal{N} (\mathbf{\Lambda}^{0},\mathbf{\sigma_\Lambda^2(M^*)}), \: \mathbf{\sigma_{\Lambda}(M^*) }\sim \mathcal{U}[0, \sigma_{\Lambda}(M^*)].
\end{align}

To inject realistic observational uncertainty into the mock data, we assign maximum noise values to each observable as follows.
For mass, we use a maximum uncertainty of $\sigma_{ M} = 0.1\,M_{\odot}$, which corresponds to approximately 5\% of a typical $2\,M_\odot$ NS. This is a conservative choice, consistent with the reported uncertainties \cite{miller2021radius}.
For the radius, we adopt a maximum uncertainty of $\sigma_R =0.3 $ km. While this may appear optimistic relative to current observational constraints, it is aligned with future observational prospects—including those from next-generation missions—which aim to reduce radius uncertainties to below ${\sim}200$ meters \cite{huxford2024accuracy,chatziioannou2022uncertainty}.
 We note that the adopted maximal deviations play a decisive role in the results: increasing the dispersions substantially broadens the posteriors and enhances parameter degeneracies, while smaller deviations—such as those expected from next-generation X-ray and GW observations—enable sharper constraints and improved physical interpretability.

To account for the strong mass dependence of tidal deformability $\Lambda$, we take as reference the uncertainty bands provided in the supplementary material of \cite{abbott2018gw170817} present in \cite{LIGO_supplemental} , focusing on the scenario that includes mass constraints. 

Specifically, from the 50\% CI shown for the low-spin case, we derived an effective standard deviation $\sigma_\Lambda(M)$. Under the assumption that the posterior distribution for $\Lambda$ at a fixed mass is approximately Gaussian, the standard deviation is related to the 50\% CI width as $\sigma = \text{CI Width} / 1.349$, since for a normal distribution the interquantile range (i.e., the 50\% interval) is approximately $2 \times 0.6745 \sigma$.

We model the standard deviation of tidal deformability as a mass-dependent function, using a quadratic fit in the logarithmic scale of $\Lambda$:
\begin{equation}
    \hat{\sigma}(M)= 121483.4 \times e^{-5M + 0.37M^2}\,,
\end{equation}
where mass $M$ is expressed in $M_\odot$, and $\hat{\sigma}(1.4\,M_{\odot})=223.8$. This formulation enables us to use a dataset-independent, yet physically motivated noise model that varies smoothly with mass.
During sampling, we ensure that all noise-perturbed values respect physical bounds: $[1,M_{\text{max}}(\text{EoS})]\,M_\odot$ and $ \Lambda >0$. Additionally, we apply a base-10 logarithmic transformation (i.e., $\log_{10}$) to the tidal deformability values and standardize all vectors of observational quantities.

The four dataset variants used in this study—reflecting the presence or absence of observational noise and tidal deformability constraints—are summarized in Table~\ref{tab:Sets}. Specifically, $R_1$ and $R_2$ correspond to datasets without tidal deformability, with $R_2$ including observational noise. In contrast, $R\Lambda_1$ and $R\Lambda_2$ include tidal deformability information, with $R\Lambda_2$ additionally incorporating noise across all observables.

\begin{table}[h!]
\caption{
Overview of dataset configurations used in this work. Each configuration combines a specific noise level and whether or not tidal deformability constraints are included. A check mark (\ding{52}) indicates presence, and a cross (\ding{55}) indicates absence.
}
\centering
\setlength{\tabcolsep}{10pt}
\renewcommand{\arraystretch}{1.2}
\begin{tabular}{lcccc}
\hline\hline
Property / Set $\mathbb{X}$  & $R_1$ & $R_2$ & $R\Lambda_1$ & $R\Lambda_2$ \\
\hline
Observational noise        & \ding{55} & \ding{52} & \ding{55} & \ding{52} \\
Tidal deformability ($\Lambda$) & \ding{55} & \ding{55} & \ding{52} & \ding{52} \\
\hline\hline
\end{tabular}
\label{tab:Sets}
\end{table}

To avoid overfitting to a fixed dataset and encourage broader generalization, we opted to sample fresh training pairs ($\mathbf{M} $, $\mathbf{R} $ and $\mathbf{\Lambda} $) during each training epoch, rather than relying on a static dataset as in our previous work \cite{carvalho2024detecting,PhysRevD.108.043031}. This approach aligns with the likelihood-free, simulation-based nature of NPE.

\subsection{Generation of Mock Test Datasets} \label{sec:gm_test}

To account for the observed mass distribution of NSs, we used the dataset presented in \cite{rocha2023mass}, which includes a comprehensive table of 125 pulsar observations. However, three of these entries report only the total system mass and not the individual pulsar mass. As this study focuses on the pulsar component, only 122 observations were considered, which are shown in Fig. \ref{fig:Mass_distr}. 

To generate realistic mock test data, we retained the same three intervals of mass used during  training -- [1.0,1.4], [1.4,1.7], [1.7, $M_{\text{max}}(\text{EoS})]\,M_\odot$ -- to ensure consistency with training distribution. However, we adjusted the number of samples per interval to reflect the empirical distribution observed in \cite{rocha2023mass}, resulting in a composition of $[6,4,5]$ samples respectively. This provides a test set that is both aligned with the model’s training and representative of astrophysical observations.
The chosen intervals reflect the empirical clustering of observed NS masses and are intended to balance model exposure across low-mass, canonical, and high-mass regimes.

\begin{figure*}[!hbt]
    \centering
    \includegraphics[width=\linewidth]{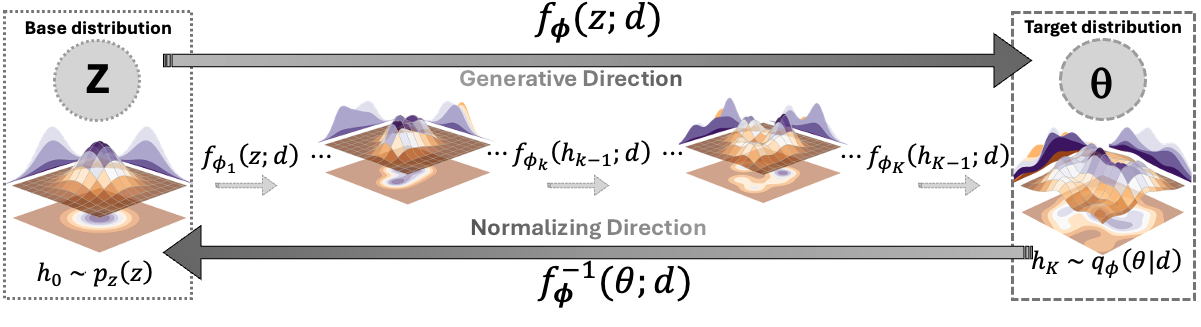}
    \caption{
Schematic illustration of CNFs. A base sample \( \boldsymbol{z} = h_0\sim \mathcal{N}(0, I) \) is transformed via a learned, invertible mapping \( f_{\boldsymbol{\phi}}(\boldsymbol{z};\boldsymbol{d}) \) across $K$ coupling layers, denoted as $f_{\phi_K}(h_{K-1};d) \circ \cdots \circ  f_{\phi_1}(z;d) $, conditioned on observations \( \boldsymbol{d} \), to yield posterior samples \( \boldsymbol{\theta} = h_K \sim q_\phi(\theta|d) \), more details can be seen in Sec. \ref{sec:cnf}. 
    The normalizing direction, \( f^{-1}_{\boldsymbol{\phi}}(\boldsymbol{\theta};\boldsymbol{d}) \), is the training direction, while the generative direction stands for the inference direction.  
    }
    \label{fig:CNF}
\end{figure*}

\section{Neural Posterior Estimation (NPE)}
\label{NPE}

NPE is a likelihood-free, SBI framework designed for amortized Bayesian inference. Its goal is to learn a parametric approximation \( q_{\boldsymbol{\phi}}(\boldsymbol{\theta} \mid \boldsymbol{d}) \) to the true posterior distribution \( p(\boldsymbol{\theta} \mid \boldsymbol{d}) \), where \( \boldsymbol{\theta} \) are the latent parameters of interest, \( \boldsymbol{d} \) denotes observed data, and \( \boldsymbol{\phi} \) are the trainable parameters of a neural network.

Unlike classical approaches such as MCMC, which can be computationally intensive and must be rerun for each new observation, NPE enables rapid inference after a single training phase, making it particularly suitable for settings where repeated inference is required. Crucially, it also retains the key advantage of Bayesian methods: principled uncertainty quantification.

NPE has found widespread applications in physics—particularly in GW astronomy, where complex, high-dimensional posteriors must be inferred without an explicit likelihood function \cite{dax2021real,Dax:2024mcn}.

\subsection{Theoretical Framework}

NPE operates by minimizing the expected Kullback–Leibler (KL) divergence between the true posterior and the learned approximation across all possible observations:
\begin{align}
L_{\text{NPE}} 
&= \mathbb{E}_{\boldsymbol{d} \sim p(\boldsymbol{d})} \left[ \text{D}_{\text{KL}}\left( p(\boldsymbol{\theta} \mid \boldsymbol{d}) \, \| \, q_{\boldsymbol{\phi}}(\boldsymbol{\theta} \mid \boldsymbol{d}) \right) \right]  \nonumber \\
&= - \mathbb{E}_{\boldsymbol{\theta} \sim p(\boldsymbol{\theta})} \, \mathbb{E}_{\boldsymbol{d} \sim p(\boldsymbol{d} \mid \boldsymbol{\theta})} \left[ \log q_{\boldsymbol{\phi}}(\boldsymbol{\theta} \mid \boldsymbol{d}) \right] \nonumber \\
&\approx -\frac{1}{M} \sum_{j=1}^M \frac{1}{N} \sum_{i=1}^N \log q_{\boldsymbol{\phi}}(\boldsymbol{\theta}^{(i)} \mid \boldsymbol{d}^{(ij)}),
\end{align}
where \( \boldsymbol{\theta}^{(i)}{\sim}p(\boldsymbol{\theta}) \) are parameter samples drawn from the prior, \( \boldsymbol{d}^{(ij)}{\sim}p(\boldsymbol{d} \mid \boldsymbol{\theta}^{(i)}) \) are simulated observations, \( N \) is the number of unique parameter samples, and \( M \) is the number of simulations per parameter.

This objective corresponds to maximum likelihood estimation on synthetic data sampled from the joint prior \( p(\boldsymbol{\theta}, \boldsymbol{d}) \). In practice, these samples are generated using a forward simulator, which encodes the underlying physical model.
Once trained, the network approximates the posterior \( q_{\boldsymbol{\phi}}(\boldsymbol{\theta} \mid \boldsymbol{d}) \) for any new observation \( \boldsymbol{d} \), enabling efficient inference with uncertainty estimates.

\subsection{Conditional Normalizing Flows} \label{sec:cnf}

We model $q_{\boldsymbol{\phi}}(\boldsymbol{\theta} |\boldsymbol{d})$ using CNFs \cite{rezende2016variational,durkan2019neural}, which are invertible neural networks that transform a simple base distribution (e.g., standard Gaussian) $p_z(\boldsymbol{z})$ into a complex posterior.  A schematic overview of this approach is provided in Fig.~\ref{fig:CNF}.
The transformation is achieved through a sequence of invertible and differentiable mappings:
\begin{equation}
q_{\boldsymbol{\phi}}(\boldsymbol{\theta}|\boldsymbol{d}) = p_z\left(f^{-1}_{\boldsymbol{\phi}}(\boldsymbol{\theta}; \boldsymbol{d})\right) \left| \det \left( \frac{\partial f^{-1}_{\boldsymbol{\phi}}}{\partial \boldsymbol{\theta}} \right) \right| \;,
\end{equation}
where $f_{\boldsymbol{\phi}}: \boldsymbol{z} \rightarrow \boldsymbol{\theta}$ denotes a bijective and differentiable transformation parameterized by $\boldsymbol{\phi}$ and conditioned on the observed data $\boldsymbol{d}$. The term $\left| \det \left( \partial f^{-1}_{\boldsymbol{\phi}}/\partial \boldsymbol{\theta}\right) \right|$  is the absolute value of the determinant of the Jacobian of the inverse transformation, which ensures proper normalization of the probability density via the change-of-variables formula.

\subsection{Physics-Informed Loss}
\label{sec:phys_loss}

To ensure that our model produces physically consistent predictions, we incorporate domain-specific inductive biases into the training process through a physics-informed loss term. Specifically, we enforce the physically expected monotonicity of the pressure–density relation:  
\textit{the pressure should increase with baryonic density}, as required by causality and thermodynamic stability in NS matter. To this end, we augment the NPE loss function with a monotonicity penalty term that discourages non-physical outputs where the predicted pressure decreases with increasing density. The total loss becomes:
\begin{equation}
L = L_{\text{NPE}} + \lambda \cdot L_{\text{monotonicity}}\;,
\end{equation}
where \( \lambda > 0 \) is a hyperparameter that controls the strength of the regularization, and the penalty term is defined as:
\begin{equation}
L_{\text{monotonicity}} = \sum_{k=1}^{19} \max \left(0, \, p(n_k) - p(n_{k+1}) \right)\;.
\end{equation}
Here, \( p(n_k) \) denotes the predicted pressure at the \(k^{\text{th}} \) baryonic density point. For each adjacent pair, the penalty contributes only when the pressure decreases, thereby enforcing: $p(n_1) \leq p(n_2) \leq \cdots \leq p(n_{20})$.

The complete training objective, including the change-of-variables likelihood from the CNF and the monotonicity penalty, becomes:
\begin{multline}
L = -\mathbb{E}_{\boldsymbol{\theta} \sim p(\boldsymbol{\theta})} \, 
\mathbb{E}_{\boldsymbol{d} \sim p(\boldsymbol{d}|\boldsymbol{\theta})} \Biggl[
\log p_z(f_{\boldsymbol{\phi}}^{-1}(\boldsymbol{\theta}; \boldsymbol{d})) + \\
\log \left| \det \frac{\partial f_{\boldsymbol{\phi}}^{-1}}{\partial \boldsymbol{\theta}} \right| \Biggr] + \lambda \cdot \sum_{k=1}^{19} \max \left(0,\, p(n_k) - p(n_{k+1}) \right).\end{multline}
This framework combines the expressive power of NPE with physically grounded regularization, ensuring that model predictions remain both flexible and interpretable. By embedding physical laws into the learning objective, we promote better generalization and trustworthiness in scientific inference.
 In contrast to the soft constraint on monotonicity, we did not enforce causality ($c_s \leq 1$) via the loss function. Implementing a hard constraint on the speed of sound within the normalizing flow framework is nontrivial, as it requires a custom, invertible transformation layer that strictly bounds the output values, which can excessively restrict the model's flexibility and lead to worse performance, especially for EoS ensembles with highly oscillatory sound speed profiles.

However, the causality of the inferred EoS is ensured implicitly by our methodology. Our training priors (both polytropic and Gaussian process) are themselves rigorously constrained to be causal \textit{a priori}. Consequently, the amortized neural posterior estimator learns a conditional distribution $q_\phi(\boldsymbol{\theta} | \boldsymbol{d})$ that is inherently biased towards the physical properties of the prior, making the prediction of sound speeds greater than the speed of light extremely unlikely. 

To validate this, we computed the fraction of posterior samples violating causality across our test sets.  The median fraction of noncausal samples across all EoSs was found to be less than 0.5\%, and the maximum fraction observed for any single EOS was 1.5\%. 
 While this already represents a high degree of physical consistency, future work could explore implementing a dedicated constrained transformation layer within the normalizing flow to entirely eliminate these rare outliers.\\

The implementation details of our work are described in Appendix \ref{ssec:implementation}.


\section{Results}
\label{results}

In this section, we evaluate the performance of our trained model in reconstructing key NS properties, specifically the pressure, squared speed of sound, and trace anomaly, as functions of the baryonic density. These reconstructions are based on synthetic observational data. Our objective is to assess the model’s ability to accurately infer the relevant physical quantities while providing well-calibrated uncertainty estimates.

To ensure a fair comparison between the PT and GP models, we fix the size of the test set to 4,000 EoSs for both datasets. These EoSs are randomly sampled from the original test set to allow consistent statistical evaluation across models.
Full validation diagnostics, including coverage and credible interval calibration, are reported in Appendix \ref{appendix:coverage}.
\subsection{Parameter Representation and Posterior Sampling}\label{sec.PS_}
We denote the fixed grid of baryon densities as a vector:

\begin{equation}
    \mathbf{n} = [n_1, n_2, \dots, n_{20}],
\end{equation}
and define the vector of physical quantities to be inferred as:

\begin{equation}
\boldsymbol{\theta} =  
    \begin{cases}
    \boldsymbol{p} = [p(n_1), p(n_2), \cdots, p(n_{20})], \\[5pt]
    \boldsymbol{c_s^2} = [c_s^2(n_1), c_s^2(n_2), \cdots, c_s^2(n_{20})],\\[5pt]
    \boldsymbol{\Delta} = [\Delta(n_1), \Delta(n_2), \cdots, \Delta(n_{20})],
\end{cases}
\end{equation}
where \(\boldsymbol{n} = [n_1, \dots, n_{20}]\) represents a fixed grid of baryon densities, and \(\boldsymbol{\theta}\) represents either the pressure, the squared speed of sound, or the trace anomaly evaluated at those points.

Our goal is to approximate the true posterior distribution \(p(\boldsymbol{\theta} \mid \boldsymbol{d})\), where \(\boldsymbol{d}\) denotes the set of observational constraints (e.g., masses, radii, tidal deformabilities). To do this, we train a CNF to learn an invertible mapping from a simple base distribution to the target parameter space:
\[
f_{\boldsymbol{\phi}}: (\boldsymbol{z}, \boldsymbol{d}) \longmapsto \boldsymbol{\theta}, \quad \text{with} \quad \boldsymbol{z} \sim \mathcal{N}(\boldsymbol{0}, \boldsymbol{I}),
\]
where \(f_{\boldsymbol{\phi}}\) is a diffeomorphic transformation parameterized by neural network weights \(\boldsymbol{\phi}\), and \(\boldsymbol{z}\) is a latent variable from a standard normal distribution.
To generate \(L = 1000\) approximate posterior samples for each of the $ N= 4000$ test observations, we evaluate
\begin{equation}
    \boldsymbol{z}^{(l)} \sim \mathcal{N}(\boldsymbol{0} , \boldsymbol{I})\;, \qquad
    \boldsymbol{\theta}^{(l,i)} = f_{\boldsymbol{\phi}}\left(\boldsymbol{z}^{(l)}; \boldsymbol{d^{(i)}}\right) \;, 
\end{equation}
$$l = 1, \dots, L\;, \qquad i = 1, \dots, N.$$
This results in a sample set \(\{\boldsymbol{\theta}^{(l,i)}\}_{l,i=1}^{L,N }\sim q_{\boldsymbol{\phi}}(\boldsymbol{\theta} \mid \boldsymbol{d})\), which approximates the true posterior.

\begin{figure}[!hbt]
    \centering
    \includegraphics[width=0.9\linewidth]{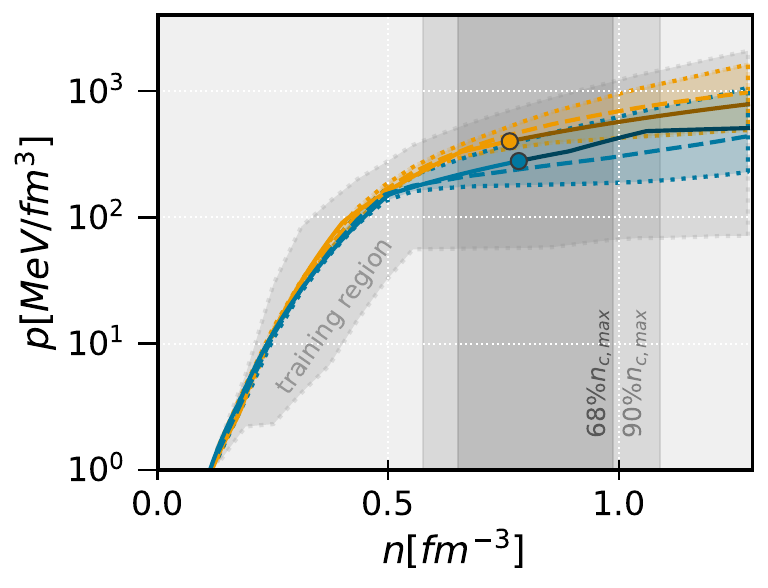}\\
     \includegraphics[width=0.9\linewidth]{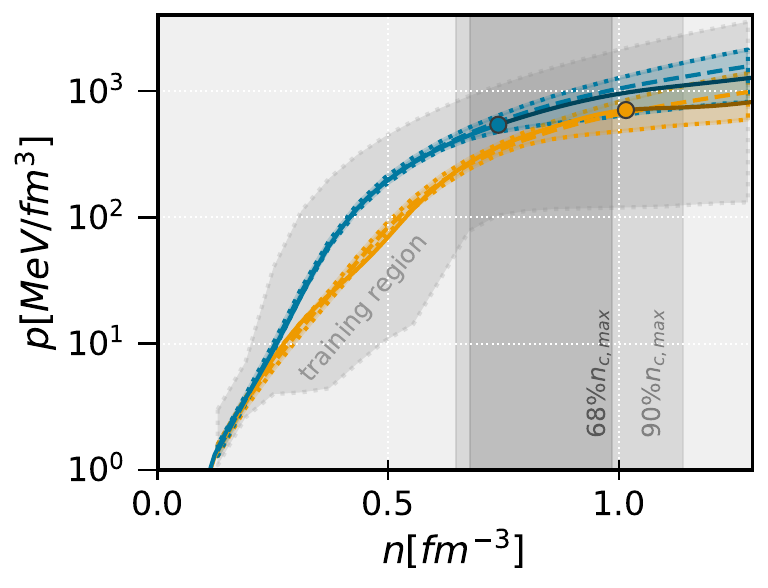}
    \caption{Reconstruction of two PT EoSs (top) and GP EoSs (bottom) from the $R\Lambda_2$ dataset. Solid lines: ground truth; shaded regions: 90\% CIs; dots: $n_{c,\max}$; vertical bands: 68\% and 90\% CI for $n_{c,\max}$ distribution.}
    \label{fig:EoS_2}
\end{figure}

\subsection{Pressure Inference on Synthetic Datasets}
\label{sec:pressure_inference}

We assess the inference performance of the model across four synthetic datasets, generated under varying noise levels and prior assumptions, see Table~\ref{tab:Sets}. These include PT and GP priors, each evaluated with and without tidal deformability and noise.

For each observation \(\boldsymbol{d}\), we sample posterior pressure vectors \(\boldsymbol{p} = [p(n_1), \dots, p(n_{20})]\) using the trained CNF, from which we compute the median and 90\% credible intervals (CIs) at each density point \(n_k\). Figure~\ref{fig:EoS_2}  shows reconstructions for two representative EoSs in the $R\Lambda_2$ set for PT and GP models, respectively. In both cases, the predicted CIs accurately enclose the true EoS up to the maximum central density \(n_{c,\max}\), with increasing uncertainty beyond this range. Vertical shaded bands indicate the distribution of \(n_{c,\max}\) values across the full sets.

To quantitatively assess the model’s accuracy, we compute the relative residual, $\text{RelRes}$, over all test-set EoSs, defined as: 
\begin{equation}\label{eq:RREs}
\text{RelRes}^{(i)}(n) = \text{Med}_l \left[ \frac{X_p^{(i,l)}(n) - X_T^{(i)}(n)}{X_T^{(i)}(n)} \right] \times 100 \;,
\end{equation}
where $\text{Med}$ denotes median, $X_p^{(i,l)}(n)$ is the predicted pressure for the $l$-th posterior sample of the $i$-th EoS at density $n$, and $X_T^{(i)}(n)$ is the corresponding true value. The model's overall performance across all EoSs is then summarized using the median absolute relative residual,  \( |\overline{\text{RelRes}}| \), at each density \( n \in \mathbf{n} \):

\begin{equation} \label{reselid}
|\overline{\text{RelRes}}|= \text{Med}_i \left| \text{RelRes}^{(i)} (n) \right|\;.
\end{equation}

\begin{figure}[t]
    \centering
\includegraphics[width=0.9\linewidth]{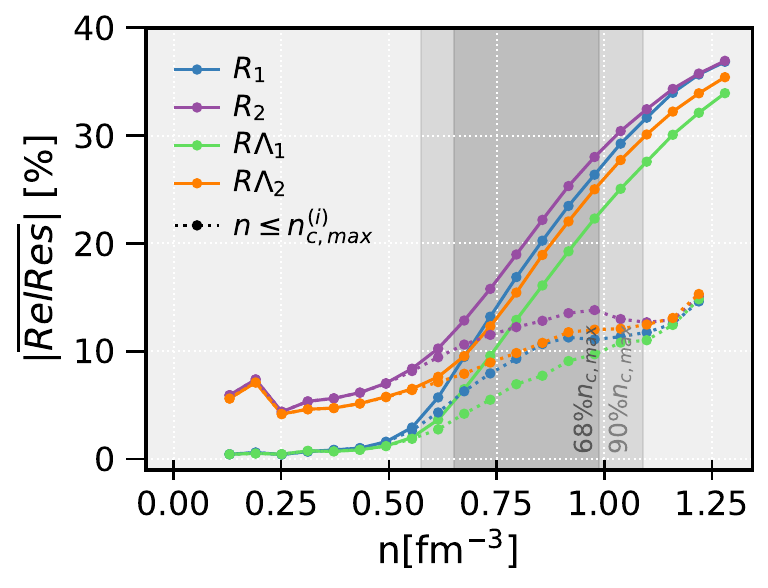}\\
    \includegraphics[width=0.9\linewidth]{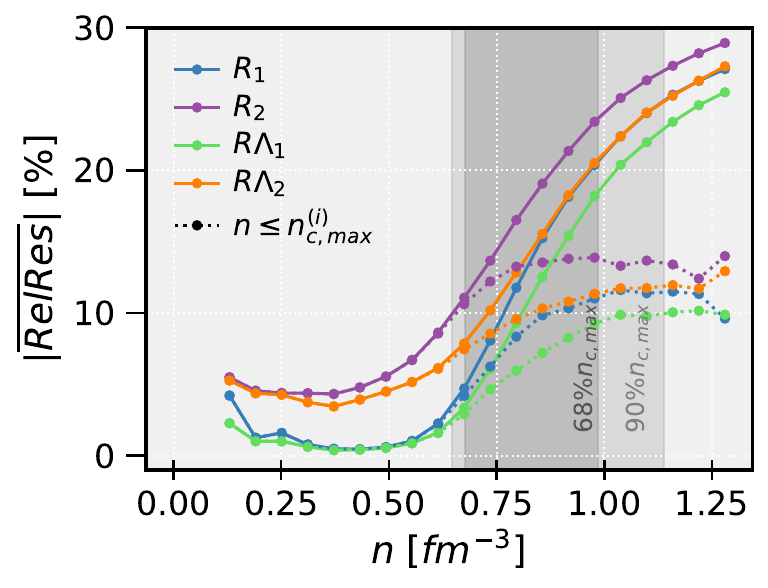}       
    \caption{ Median absolute relative reconstruction error \( |\overline{\text{RelRes}}| \) (Eq.~\ref{reselid}) across all four datasets for the PT (top) and GP (bottom) models.  Solid lines represent the error computed over the full density range, while dotted lines correspond to the truncated version $|\overline{\text{RelRes}}|_{\leq n_{c,max}}$ (Eq.~\ref{reselid_tr}) evaluated only up to the maximum central density \( n_{c,\mathrm{max}} \) of each EoS.}
    \label{fig:poly_cove}
\end{figure}

\begin{figure}[t]
    \centering
        \includegraphics[width=0.9\linewidth]{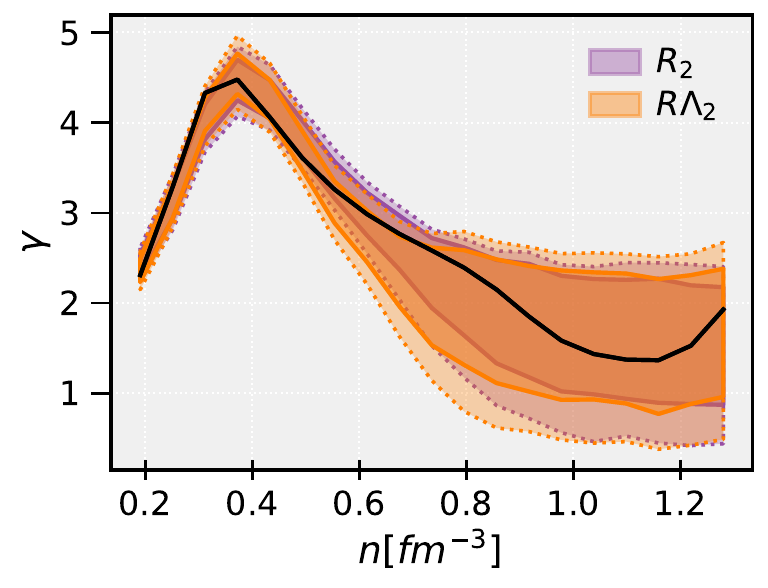}\\
                        \includegraphics[width=0.9\linewidth]{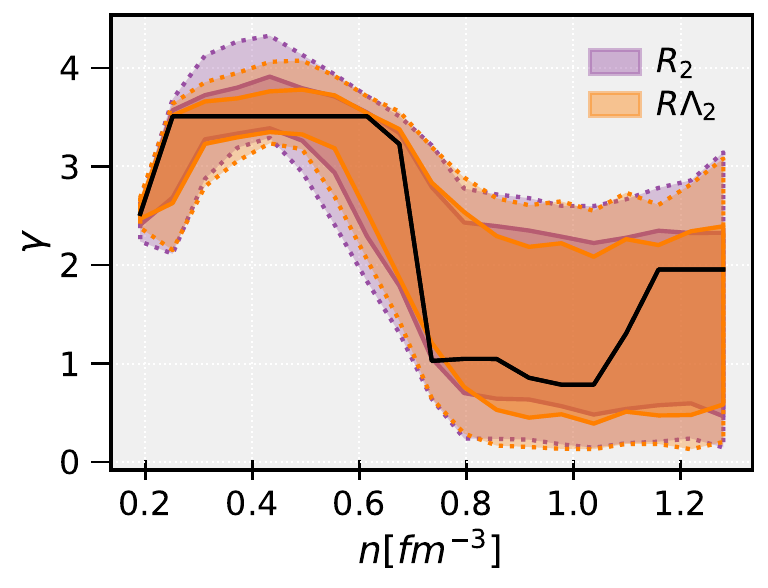}
    \caption{Reconstructed polytropic index $\gamma(n)$ for a representative EoS. Top panel: the PT model; bottom panel: the GP model. 
    The black curve shows the ''ground truth'' value, while the shaded bands represent the 68\% (solid) and 90\% (dotted) CIs for datasets $R_2$ (purple) and $R\Lambda_2$ (orange).}
    \label{fig:poly_pol}
\end{figure}

Figure~\ref{fig:poly_cove} shows this metric evaluated across all four datasets for the PT (top) and GP (bottom) models. The solid lines correspond to the full density range, while the dotted lines represent the same quantity truncated at the maximum central density $n_{c,max}^{(i)}$ of each $i$-th EoS:
\begin{equation} \label{reselid_tr}
|\overline{\text{RelRes}}|_{\leq n_{c,max}}= \text{Med}_i \left| \text{RelRes}^{(i)} (n) \right|\quad\text{for}\quad n \leq n_{c,max}^{(i)}  \; .
\end{equation}
This truncated version reflects the model’s performance only in the physically accessible region for each star and illustrates how uncertainty and bias grow beyond that regime.

At lower baryon densities, the relative error remains small across all models and increases significantly at higher densities. The behavior at high densities is mitigated when excluding the post-$n_{c,max}$ region, which leads to a marked reduction in the median residual, often from 40\% down to below 20\%. This suggests that the dominant contribution to reconstruction error at high densities, up to $1.28$~fm$^{-3}$, arises from regions beyond the maximum central density, where the model’s predictive power diminishes due to a lack of observational support.

 Notably, the dotted lines representing $|\overline{\text{RelRes}}|_{\leq n_{c,max}}$ exhibit a distinct change in behavior at the highest baryonic densities, because beyond the lower limit of the band identifying the $n_{c,max}$, the number of EoSs contributing to defining the residuals decrease and the EoSs that are not constrained by observations due to a small $n_{c,max}$ are removed. In particular, beyond the 90\% CI of $n_{c,max}$, only a very small fraction of EoSs contribute to the statistics: for the PT dataset, just 195 out of 4000 ($\sim$4.9\%) exceed this threshold, while for the GP dataset, the number is 147 EoSs ($\sim$3.7\%).

As expected, models trained without noise consistently exhibit lower residuals at low densities. Inclusion of tidal deformability information significantly reduces the error at higher densities, indicating where it provides additional constraints on the EoS.
Errors systematically increase from $R_1$ to $R_2$ and from $R\Lambda_1$ to $R\Lambda_2$, consistent with the larger noise levels in the latter training datasets.  Additionally, the error tends to grow with increasing baryon density, reflecting the model’s greater difficulty in constraining the high-density regime. Importantly, incorporating tidal deformability information (i.e. in the $R\Lambda$ datasets) leads to a notable reduction in reconstruction error, consistent with previous findings \cite{PhysRevD.110.123016,PhysRevD.108.043031}. This underscores the importance of multimessenger observations in improving constraints on the EoS.

To further quantify the effect of truncation at the maximum central density, we introduce the metric
\begin{equation}
    \eta = \frac{\sum_{n}^{20} |\overline{\mathrm{RelRes}}|_{\leq n_{c,\max}}}{\sum_{n}^{20} |\overline{\mathrm{RelRes}}|}\;,
\end{equation}
which measures the fraction of the reconstruction error arising within the physically accessible region relative to the total error across all density points. For the GP dataset we obtain values $(0.55,\,0.67,\,0.51,\,0.63)$ respectively for the ($R_1$, $R_2$, $R\Lambda_1$, $R\Lambda_2$) sets,  while for the PT dataset we find $(0.41,\,0.53,\,0.41,\,0.53)$. These results show that, once noise is included, the effect of truncation at the maximum central density becomes less pronounced (as reflected in the larger ratios), and that in the GP case the sensitivity to the maximum central density is enhanced when tidal deformability information is used.

The model’s ability to recover the polytropic index, defined as $\gamma (n) = d (\ln p(n)) / d(\ln \epsilon(n)) $ is illustrated in Fig.~\ref{fig:poly_pol} for both EoS models. The reconstructed index is shown for a representative EoS, with 68\% (solid bands) and 90\% (dotted bands) CIs for two datasets: $R_2$(purple) and $R\Lambda_2$ (orange).
The true index $\gamma(n)$ (black curve) lies entirely within the 68\% CI, confirming that both models accurately capture the derivative structure of the EoS. 
This result is particularly noteworthy given the reconstruction performed on the PT dataset, which poses additional challenges due to its discontinuities in the derivative behavior. Accurately recovering $\gamma(n)$ in such conditions highlights the model’s ability to learn and generalize the underlying physical structure, even in regimes with sharp transitions.

\begin{figure*}[t]
    \centering
        \includegraphics[width=0.9\linewidth]{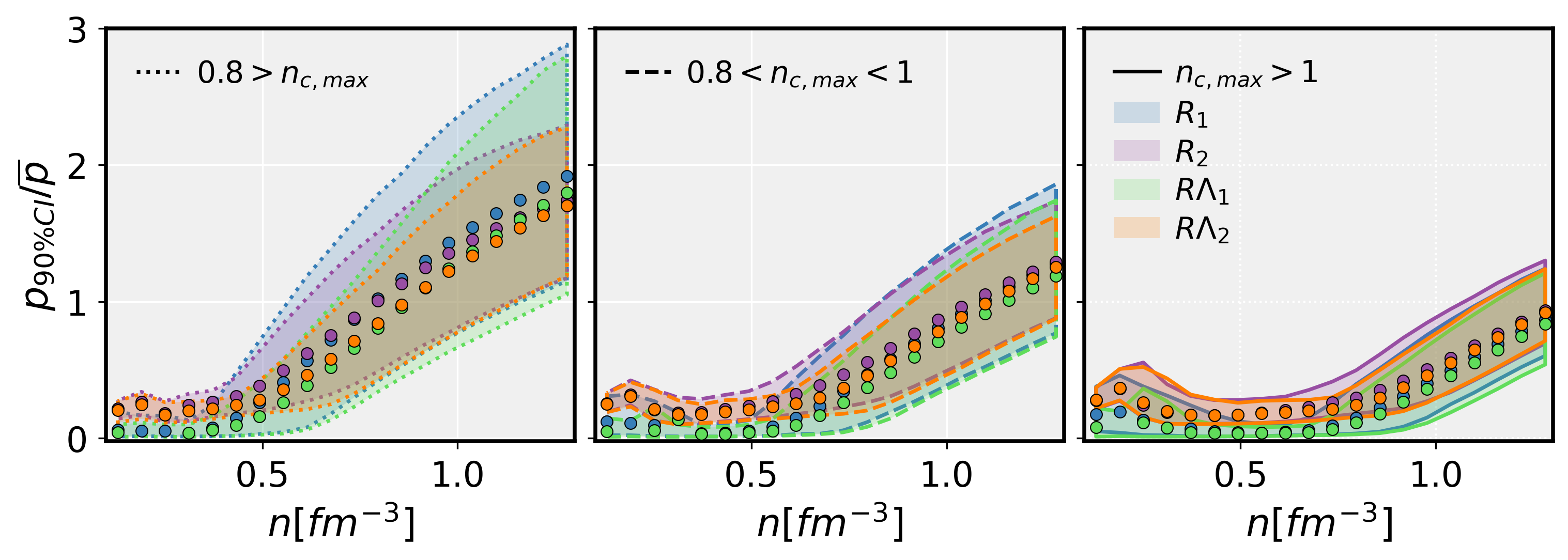}
        \includegraphics[width=0.9\linewidth]{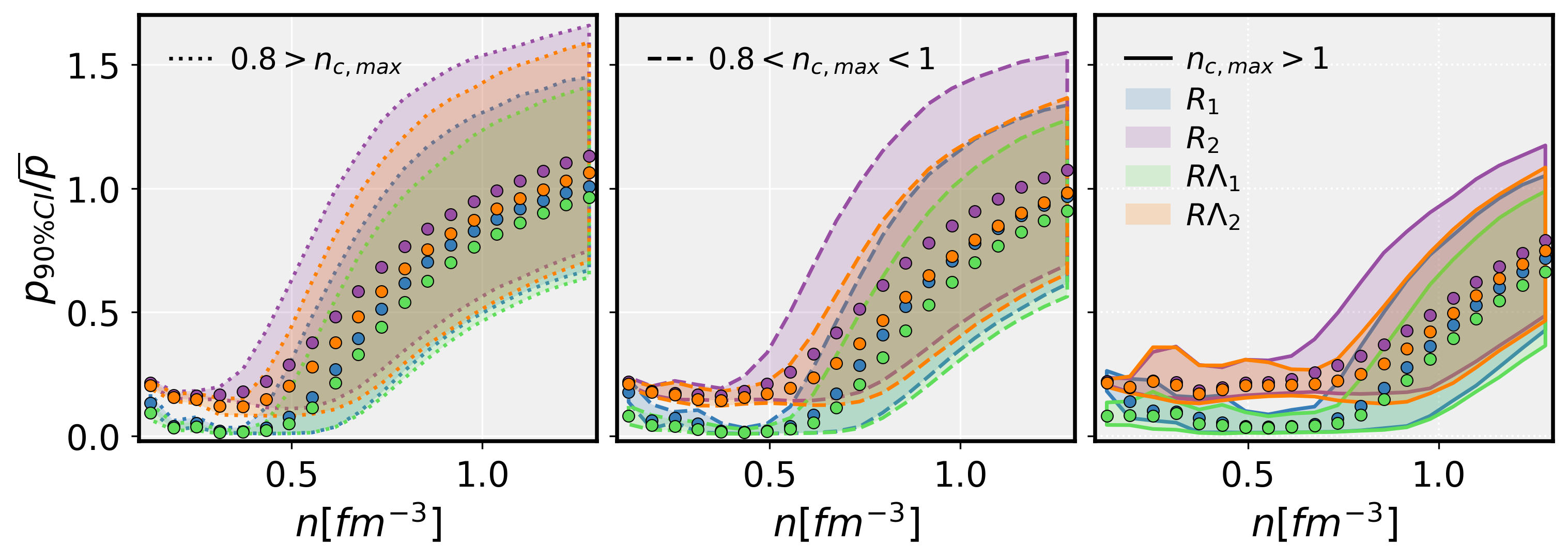}
    \caption{ 
Normalized pressure dispersion, $p_{90\%CI}/\overline p$, plotted against the baryonic density $n$ across all sets, for the PT (top) and GP (bottom)  models, divided in three intervals of $n_{c,max}$, for values smaller than 0.8 fm$^{-3}$ (left), between 0.8 fm$^{-3}$ and 1 fm$^{-3}$ (center), and greater than 1 fm$^{-3}$ (right). The shaded bands represent the 90\% CIs and the dots represent the mean.}
    \label{fig:disper_poly}
\end{figure*}

To further evaluate prediction uncertainty, we compute the normalized dispersion of the pressure predictions, defined as the ratio between the width of the 90\% CI and the median pressure value, ${p_{90\%CI}}/{\overline p}$. This quantity is plotted in Fig.~\ref{fig:disper_poly} as a function of the baryonic density, $n$, with test samples grouped by their maximum central density, $n_{c,max}$: below  0.8 fm$^{-3}$ (left), between 0.8 fm$^{-3}$ and 1 fm$^{-3}$ (center), and above 1 fm$^{-3}$ (right).
Each colored band represents a different model, with the black-edged dots denoting the mean dispersion and the shaded regions showing the 90\% CIs across all test samples.

As expected, dispersion generally increases with baryon density, reflecting growing uncertainty in the model in the high-density regime. Furthermore, models with a higher $n_{c,max}$ tend to exhibit flatter dispersion curves that begin to rise at larger $n$ values. This is consistent with these models containing more information in denser regions. In contrast, for EoSs with lower $n_{c,max}$, the dispersion increases more sharply and at lower densities, indicating that the model becomes uncertain earlier in these cases. This behavior confirms that the model correctly learns to predict confidence based on the maximum central density regime to which it was exposed during training.

\subsection{Speed of Sound and Trace Anomaly}
\label{sec:speed_sound}

We now extend our analysis to the derivative quantities of the EoS: the squared speed of sound $c^{2}_{s}(n)$ and the trace anomaly $\Delta(n)$, as defined in Section \ref{sec:ge_MD}.

\begin{figure}[h]
    \centering
        \includegraphics[width=0.9\linewidth]{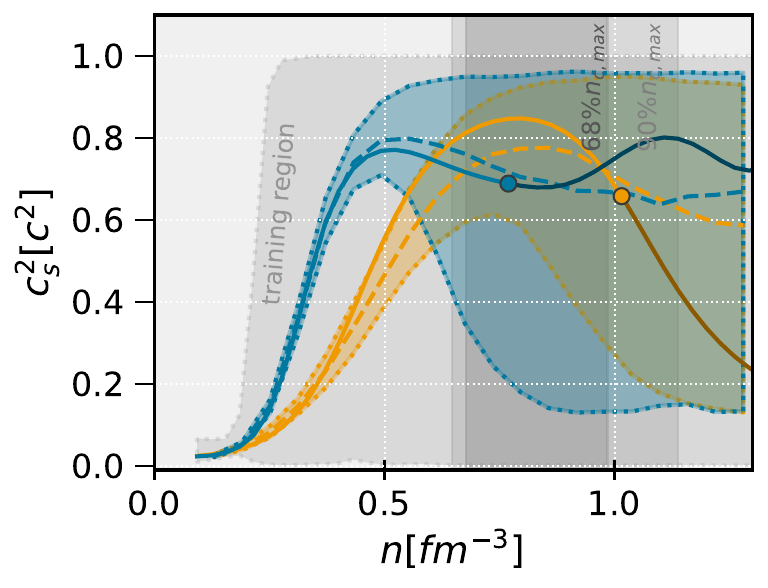}\\
        \includegraphics[width=0.9\linewidth]{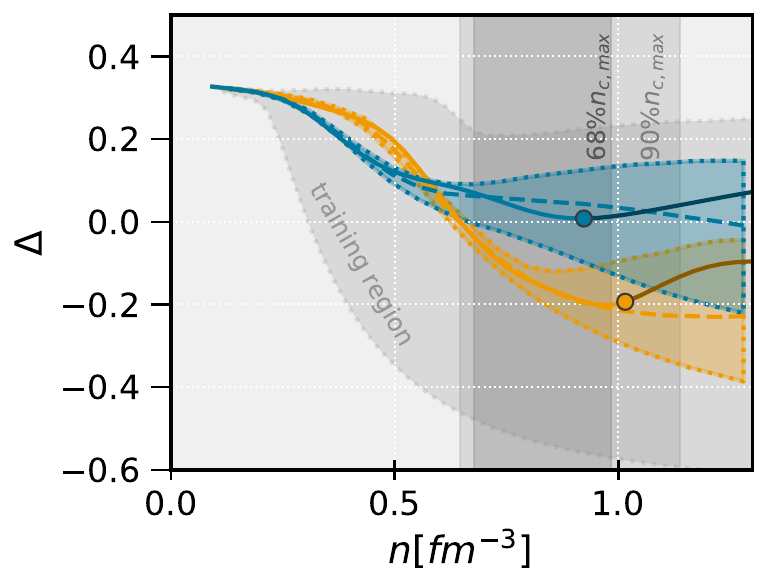}
    \caption{ Reconstruction of \( c_s^2(n) \) (top) and  trace anomaly \( \Delta(n) \) (bottom), for two representative EoSs from the GP-trained model ($R\Lambda_2$). Solid lines: ground truth; dashed lines: posterior medians; shaded areas: 90\% CIs; dots mark $n_{c,\max}$. Vertical gray bands show the 68\% (dark) and 90\% (light) CIs of $n_{c,\max}$ across the full set.}
    \label{fig:EoS_vs}
\end{figure}

Figure~\ref{fig:EoS_vs} shows the reconstruction of \( c_s^2(n) \) (top) and \( \Delta(n) \) (bottom) for two representative EoSs from the GP-trained model evaluated on the $R\Lambda_2$ dataset.

The model captures the behavior of \( c_s^2 \) well up to the first local maximum, after which the predictive uncertainty increases sharply. This is attributed to the highly oscillatory nature of the speed of sound at high densities, making it difficult for the model to generalize. In contrast, the trace anomaly is well reconstructed up to the maximum central density \( n_{c,\max} \), beyond which the model’s confidence drops significantly, as expected in regions lacking observational support.

\begin{figure}[h]
    \centering
        \includegraphics[width=0.9\linewidth]{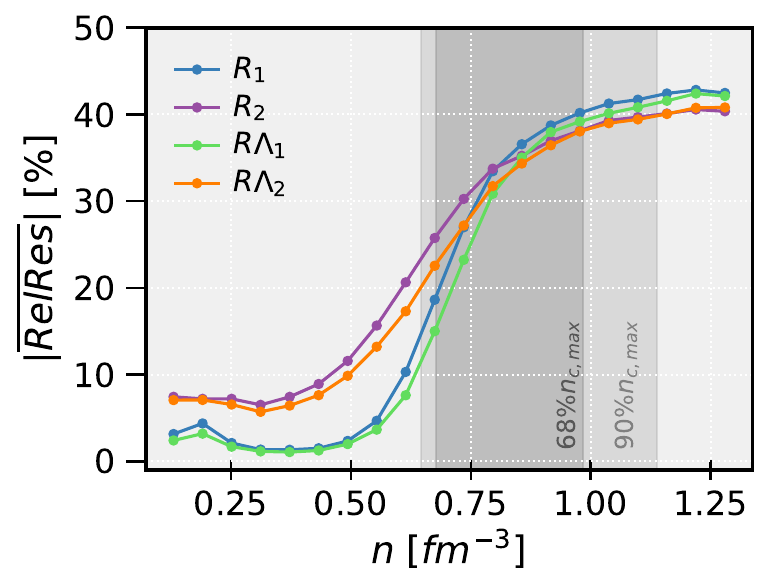}\\
        \includegraphics[width=0.9\linewidth]{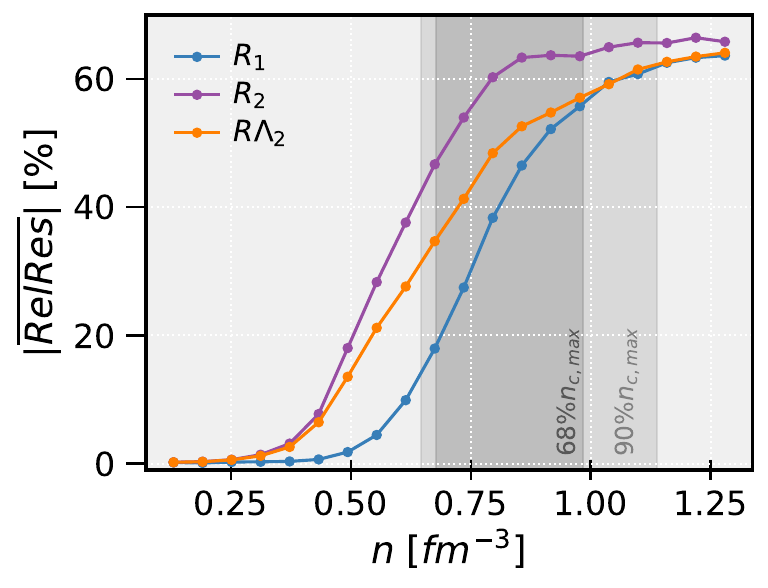}
    \caption{Median absolute relative reconstruction error \( |\overline{\text{RelRes}}| \) for \( c_s^2(n) \) across all four datasets (top) and \( \Delta(n) \) for three datasets (bottom). Vertical gray bands show the 68\% (dark) and 90\% (light) CIs of $n_{c,\max}$ across the full set.}
    \label{fig:RE_vs_}
\end{figure}

Figure~\ref{fig:RE_vs_} presents the median relative reconstruction error \( |\overline{\text{RelRes}}| \), Eq. \ref{reselid}, for \( c_s^2 \) (top) and \( \Delta \) (bottom) across all datasets. The error for \( c_s^2 \) increases with density and is only modestly reduced when tidal deformability is included, in contrast to pressure. This reflects the weaker connection between tidal deformability and local stiffness. Additionally, noise has a strong impact at low densities, where noise-free datasets yield lower errors.

For the trace anomaly, the relative error similarly grows with density, driven by the increased difficulty of reconstructing this quantity far from the data-supported regime. Nonetheless, the model retains good qualitative performance.

\begin{figure*}[!hbt]
    \centering
            \includegraphics[width=0.9\linewidth]{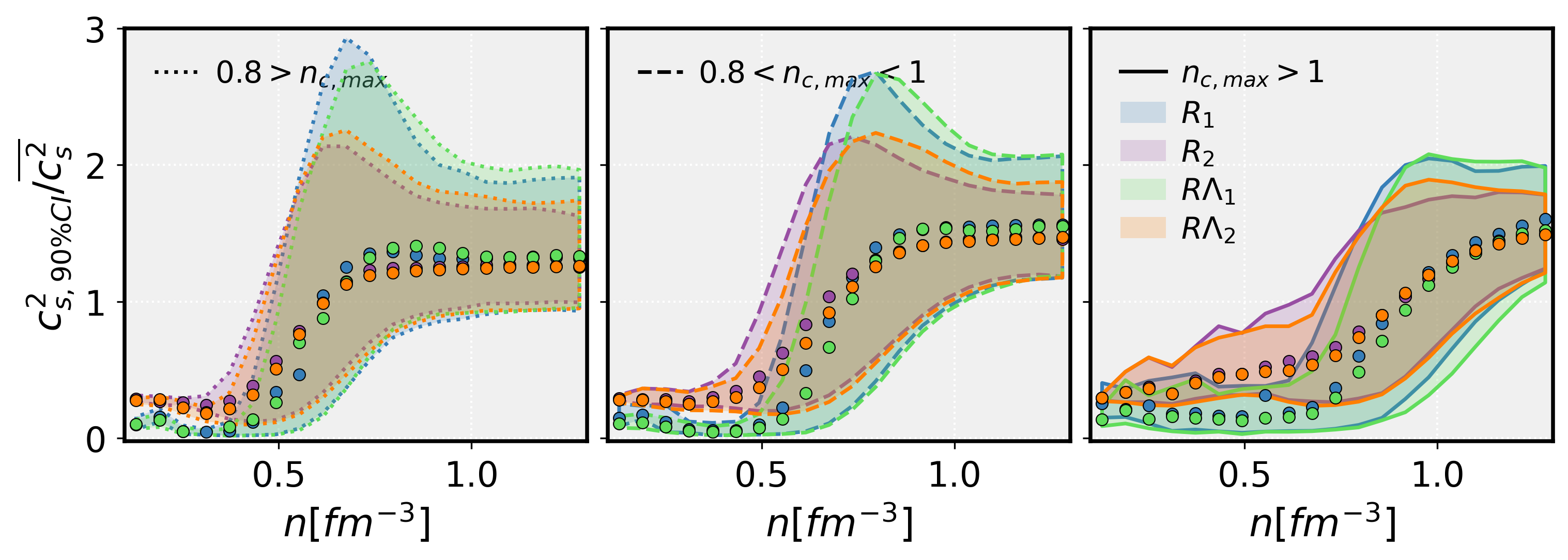}
        \includegraphics[width=0.9\linewidth]{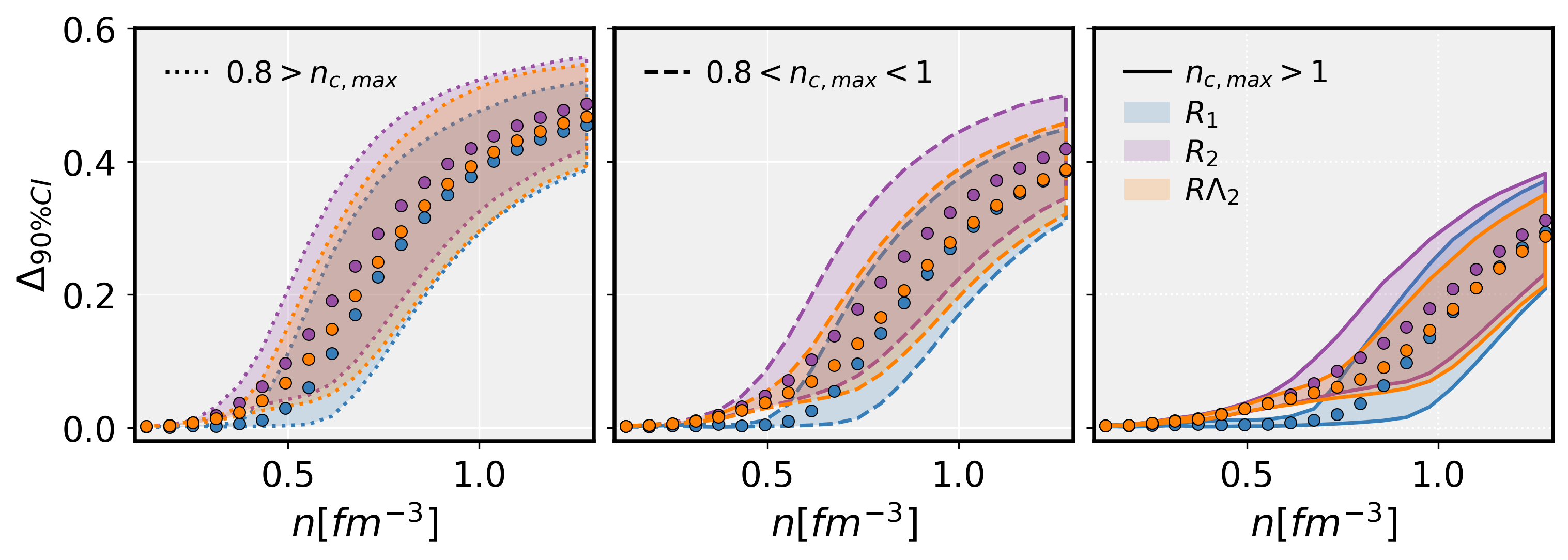}
    \caption{Normalized predictive dispersion for \( c_s^2(n) \) (top) and \( \Delta(n) \) (bottom), grouped by maximum central density. Each plot corresponds to a different interval of \( n_{c,\max} \): below 0.8~fm$^{-3}$ (left), 0.8–1.0~fm$^{-3}$ (center), and above 1.0~fm$^{-3}$ (right). Shaded bands show 90\% CIs and the dots the mean}.
    \label{fig:disp_tra}
\end{figure*}

To better understand the model's uncertainty, we plot the normalized predictive dispersion in Fig.~\ref{fig:disp_tra}. For \( c_s^2 \), we compute \( c_{s,90\%CI}^2 / \overline{c_s^2} \); for \( \Delta \), we plot the absolute 90\% CI width, since the quantity spans both positive and negative values and cannot be normalized by the median.

For the speed of sound, we observe a sharp peak near \( n \sim n_{c,\max} \), especially in the subset with lower maximum central densities. This localized rise in uncertainty indicates that \( c_s^2 \) dispersion could serve as an indirect indicator for \( n_{c,\max} \). In noise-free settings, this effect is particularly pronounced, as the model more confidently captures the underlying structure until the observational support ends.

For the trace anomaly, a similar pattern emerges: dispersion rises with density and is delayed for EoSs with larger \( n_{c,\max} \), consistent with the expectation that the model remains confident up to the limits of the data-driven regime.

Figures~\ref{fig:disper_poly} and \ref{fig:disp_tra} suggest a possible correlation between the maximum central baryon density inside NSs ($n_{c,\max}$) and the slope of the normalized predictive dispersion. To explore this further, we investigated the relationship between $n_{c,\max}$ and the density point at which the dispersion exhibits the steepest growth in Appendix \ref{sec:AP.corre}. The results, presented there in Fig.~\ref{fig:correla}, show a weak correlation for pressure in the PT dataset. In contrast, the GP dataset displays a stronger correlation—particularly for the squared speed of sound and the trace anomaly—where Pearson coefficients exceed 0.7. Notably, this correlation persists even in the presence of observational noise.

\subsection{Cross-Dataset Generalization}
\label{sec:cross_dataset}

\begin{figure*}[!hbt]
    \centering
        \includegraphics[width=0.33\linewidth]{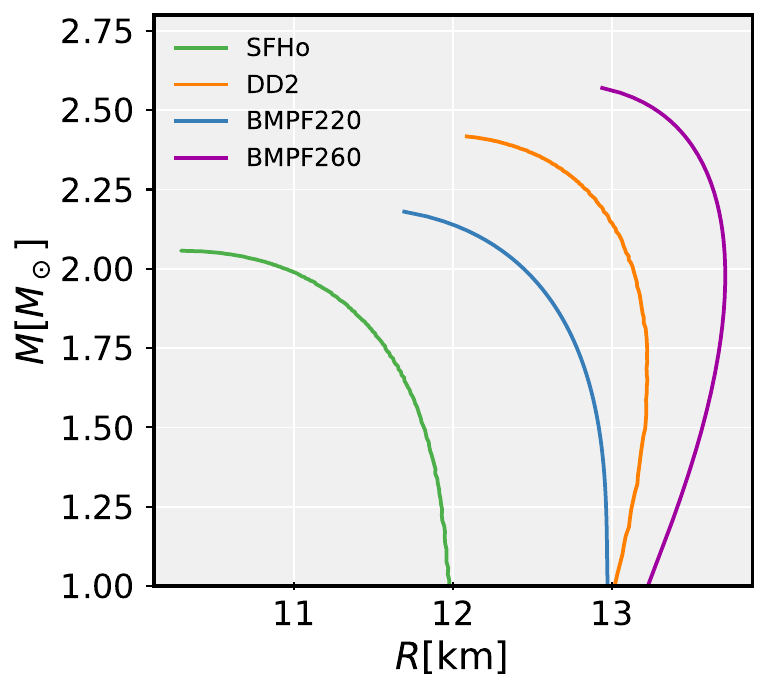}
                \includegraphics[width=0.33\linewidth]{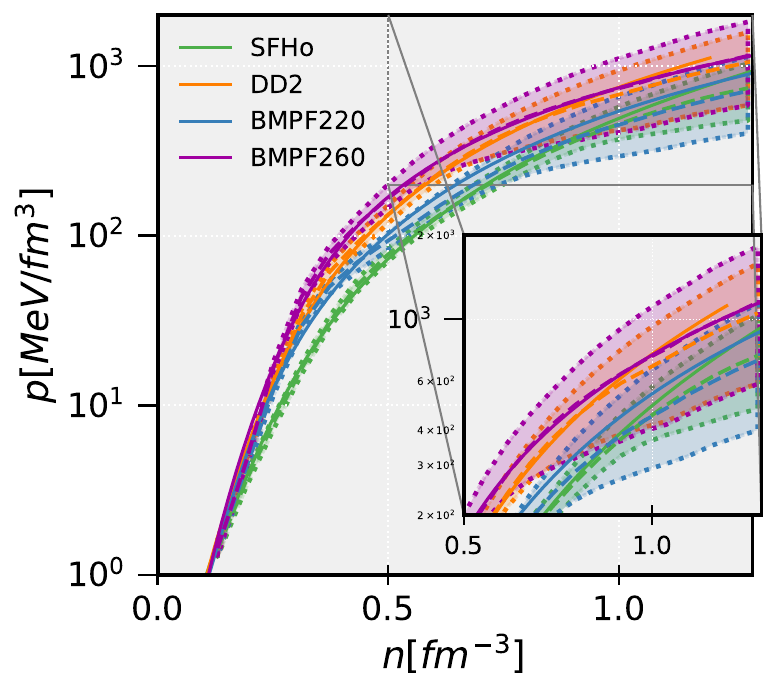}
        \includegraphics[width=0.33\linewidth]{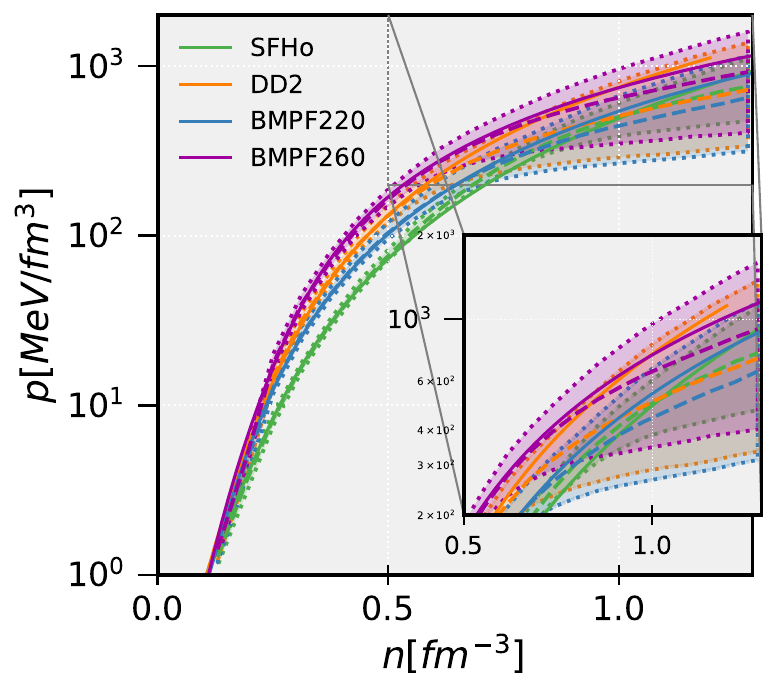}
    \caption{
    Left panel: Mass–radius relations for the four test EoSs not seen during training.  
    Middle panel: Pressure predictions from the model trained on the GP dataset.  
    Right panel: Pressure predictions from the model trained on the PT dataset.  
    Solid lines represent the true EoSs, while shaded regions show 90\% CIs from the models and dashed lines represent the median. The insets in the pressure plots, restrict the density range to [0.5:1.28]~fm$^{-3}$ and the pressure range to [2$\times$10$^2$:2$\times$ 10$^3$]~MeV/fm$^{3}$}
    \label{fig:MR_new}
\end{figure*}

To assess the robustness and generalization ability of the $R_2$-trained models for pressure, we tested both the PT- and GP-based architectures on four EoS models not seen during training. Two of these are phenomenological EoSs—BMPF220 (blue) and BMPF260 (purple)—introduced in the supplementary material of \cite{malik2023spanning}. The third is the SFHo model (green) from \cite{steiner2013core}, and the fourth is the DD2 EoS (orange), described in \cite{fortin2016neutron}.
These four EoSs are based on physically motivated parameterizations that differ significantly from the agnostic PT and GP families used during training. Their $M(R)$ curves are also highly distinct, as shown in the left plot of Fig.~\ref{fig:MR_new}, making them ideal for evaluating the extrapolation capability of the trained models.

The main aim of this test is to establish whether models trained on agnostic datasets without any explicit nuclear microphysics can still provide meaningful reconstructions when faced with realistic, phenomenological EoSs. If successful, this would demonstrate the potential of SBI to support our understanding of dense matter inside NS cores.

Figure~\ref{fig:MR_new} demonstrates that both models are indeed capable of producing reasonable predictions even when applied to very distinct EoSs. The GP-trained model ( middle  plot) shows slightly better coverage of the true EoSs, while the PT-trained model (right plot) remains competitive despite being trained on a more constrained family of parametrizations.

\begin{figure}[!hbt]
    \centering
        \includegraphics[width=0.9\linewidth]{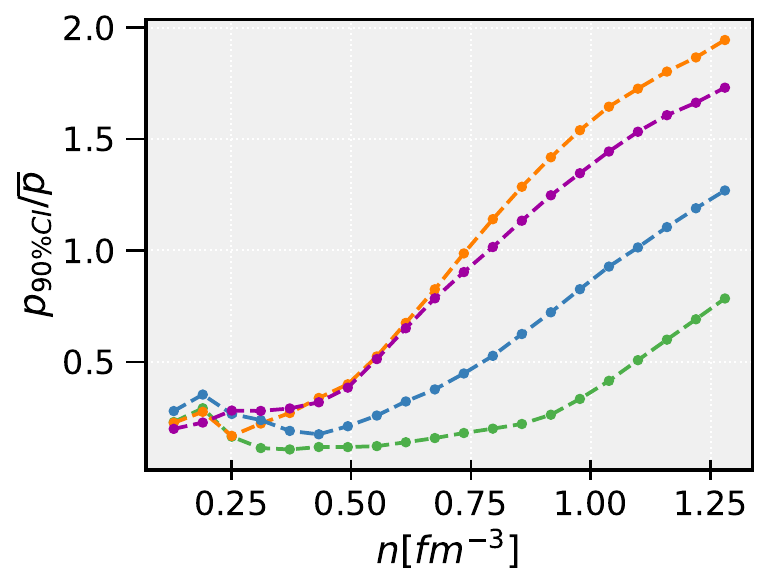}\\
                \includegraphics[width=0.9\linewidth]{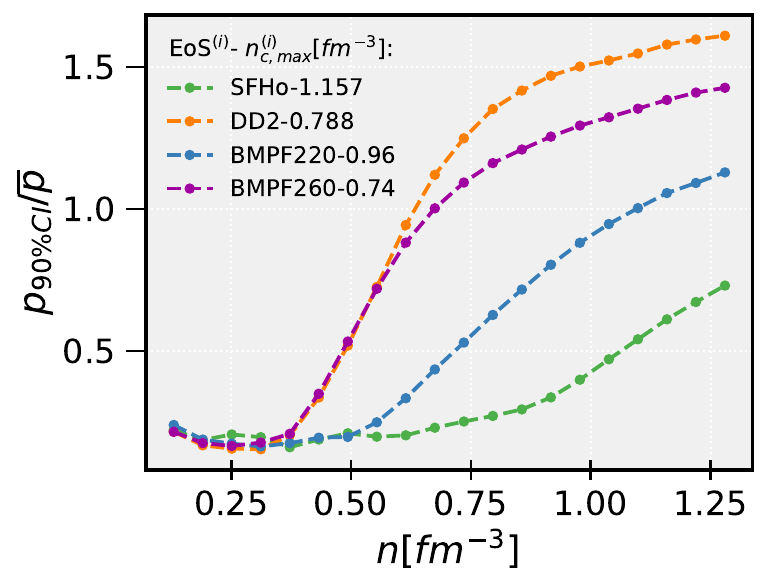}
    \caption{Normalized pressure dispersion, $p_{90\%CI}/\overline p$,  plotted as a function of baryonic density $n$ for four representative EoSs, with their respective maximum central densities indicated in the legend. The top plot corresponds to the PT dataset, and the bottom plot to the GP dataset.}
    \label{fig:p90p_new}
\end{figure}

From Fig.~\ref{fig:p90p_new}, we observe three distinct regions in the predictive dispersion, corresponding to different ranges of maximum central density. These bands reflect the model’s ability to infer the region associated with the maximum central density. EoSs with higher $n_{c,max}$ exhibit flatter dispersion curves that begin to rise more gradually, with the increase occurring closer to  $n \sim 1$ fm$^{-3}$. In contrast, EoSs with lower $n_{c,max}$ display much steeper curves, with dispersion increasing significantly at lower densities. This behavior suggests that the model implicitly captures information related to the central density, despite not being explicitly trained to predict it.

\begin{figure}[!hbt]
    \centering
    \includegraphics[width=0.9\linewidth]{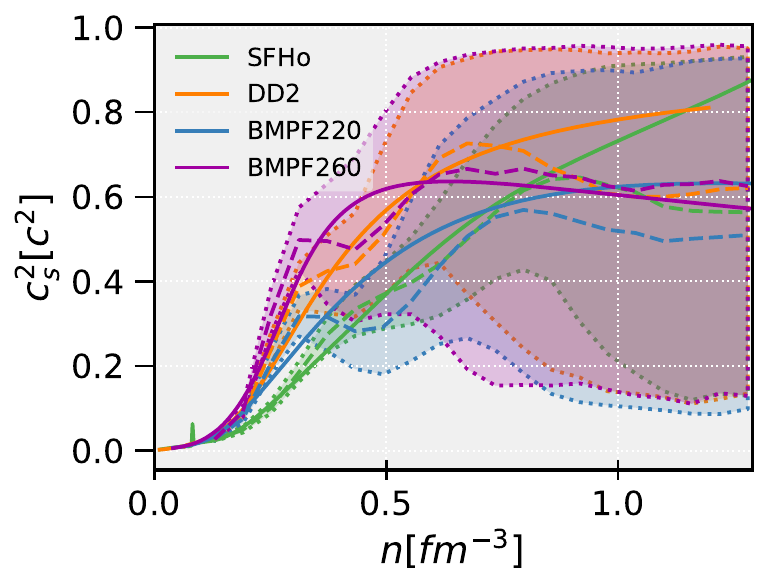}\\
        \includegraphics[width=0.9\linewidth]{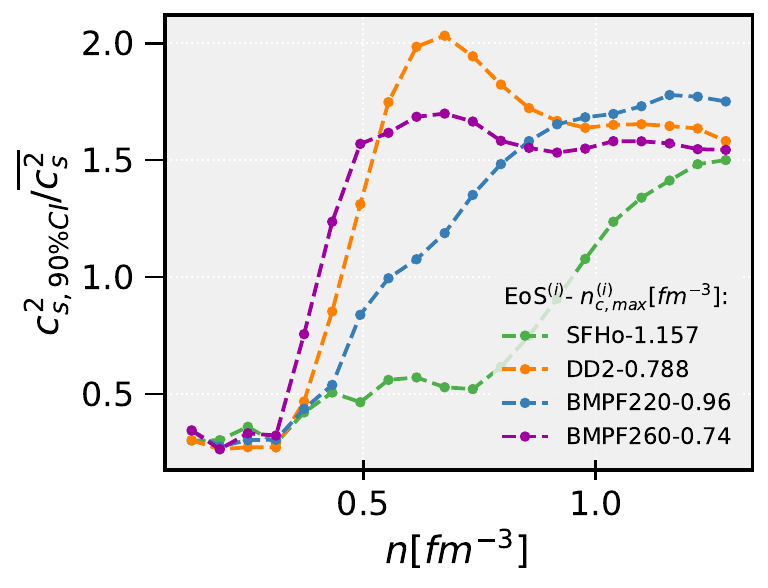}
    \caption{Speed of sound squared predictions  (top) and  normalized speed of sound squared dispersion (bottom)  plotted as a function of baryonic density $n$ for four representative previously unseen EoSs, with their respective maximum central densities indicated in the legend.
Top panel:  Solid lines represent the true EoSs and the dashed lines the medians, and the shaded regions show 90\% CIs from the models.}  
    \label{fig:vs90vs_new}
\end{figure}

In Fig.~\ref{fig:vs90vs_new}, we present the model predictions for the squared speed of sound across the four test EoSs. While the predicted median curves exhibit noticeable oscillations, the true speed of sound remains consistently within the 90\% credible intervals, demonstrating reliable uncertainty coverage.

The bottom plot shows the normalized dispersion, where we again observe that models corresponding to lower $n_{c,max}$ values display a sharper increase in uncertainty at lower densities—mirroring the behavior seen in the pressure predictions. Interestingly, the peak in dispersion typically occurs slightly before reaching the maximum central density, which is consistent with the fact that the model is not explicitly conditioned on the maximum mass. For EoSs with higher $n_{c,max}$,  however, the dispersion does not decline as sharply as it does in the pressure case, suggesting a different sensitivity profile for the speed of sound.

\begin{figure}[!hbt]
    \centering
        \includegraphics[width=0.9\linewidth]{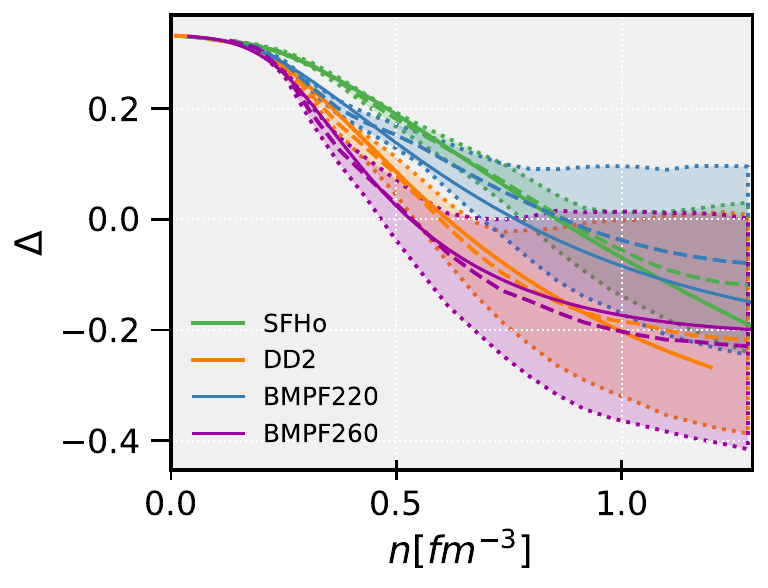} \\
        \includegraphics[width=0.88\linewidth]{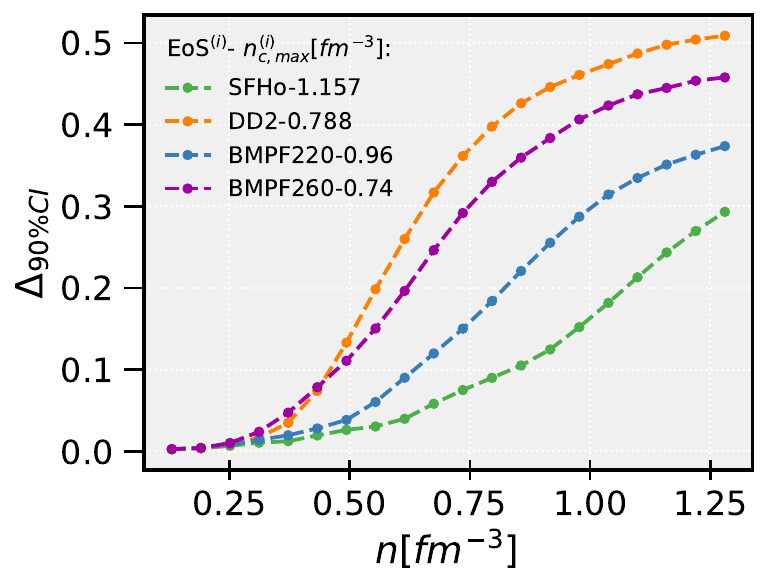}
    \caption{Top: Trace anomaly predictions for the four test EoSs. Solid lines represent the true EoSs, the dashed lines the medians and the shaded regions show 90\% CIs from the models.  Bottom: Trace anomaly dispersion, \( \Delta_{90\%CI} \), versus baryon density, with their respective maximum central densities indicated in the legend.}
    \label{fig:tr90tr_new}
\end{figure}

Finally, in Fig.~\ref{fig:tr90tr_new}, we present the predictions for the trace anomaly \( \Delta(n) \). As with other quantities, the model captures the general shape and amplitude of \( \Delta(n) \), and the true values are consistently enclosed within the predictive bands. The lower plot confirms that dispersion again stratifies according to \( n_{c,\max} \), with the onset of uncertainty increasing with maximum central density—mirroring the behavior observed in pressure and sound speed predictions.

\vspace{1em}
\noindent

Overall, these results demonstrate that our NPE models—though agnostic to underlying microphysics—can generalize remarkably well to physically motivated EoSs. Moreover, the structure of the predictive dispersion provides indirect yet consistent information about the internal density scales probed by each EoS, offering a valuable diagnostic of model confidence and applicability.

\newpage

\section{Conclusions \label{conclusion}}

In this work, we have explored the use of NPE with CNFs to infer the EoS of NS matter from synthetic observational data. Our approach is likelihood-free, allowing us to model complex posterior distributions over physical quantities such as pressure,  squared speed of sound, and trace anomaly, given mock observations of NS masses, radii, and tidal deformabilities.

To ensure physically consistent predictions, we introduced a physics-informed penalty in the loss function to enforce monotonicity in pressure — a critical feature for real-world applicability.

Our models were trained on two distinct families of agnostic sampled EoSs: PT and GP representations for pressure, and only GP-based data for the squared speed of sound and trace anomaly. Each family generated four datasets covering different observational scenarios, including combinations of tidal deformability and injected noise. This setup allowed us to systematically probe the impact of observational precision and multimessenger constraints on EoS inference.

Our main findings are:
\begin{itemize}
    \item Reliable posterior inference: The CNF model reconstructs pressure with high accuracy and well-calibrated uncertainty, even under moderate observational noise.
    \item Role of multi-messenger data: Including tidal deformability improves constraints, particularly at higher densities, where the M–R relation alone is less informative.
    \item Higher-order features: While quantities like the squared speed of sound and trace anomaly are harder to predict due to their derivative nature, the model still captures their general behavior — especially in noise-free regimes.
    \item Predictive confidence aligned with physical density limits: A key result is that the model learns to associate increasing uncertainty with proximity to the maximum central density $n_{c,max}$. We observe a consistent trend where the onset of predictive dispersion in pressure, sound speed, and trace anomaly shifts with $n_{c,max}$. This suggests that the model has implicitly learned physical boundaries, even though it is not explicitly conditioned on the maximum mass or central density.
    \item Generalization across priors and unseen EoSs: The model generalizes well across both PT and GP priors, and performs robustly on out-of-distribution EoSs with realistic nuclear microphysics, highlighting the flexibility of the SBI framework.
\end{itemize}

These findings establish NPE with CNFs as a flexible, physically informed, and data-efficient tool for NS EoS inference — capable of robust generalization and meaningful uncertainty quantification.

Looking forward, this work lays the groundwork for applying SBI to future multimessenger astrophysics, including real GW and X-ray observations. The flexibility and scalability of the CNF-based NPE framework also make it a promising tool for high-performance computing (HPC) environments, particularly for training large models on extended datasets or incorporating more complex observational constraints.
In future work, we aim to extend this approach, applying it to real observational datasets from missions such as NICER and LIGO-Virgo-KAGRA; explore inference of additional EoS-related quantities; investigate more expressive flow architectures and hierarchical priors to better capture multimodal posteriors; and leverage HPC resources for larger-scale model ensembles and uncertainty quantification. 
While this work explores how the maximum central density influences the model's predictions for the EoS, it does not aim to directly estimate or recover this value quantitatively. This is primarily because the observable inputs (mass–radius and tidal deformability points) used to condition the model are sampled randomly, without explicitly ensuring that the highest-mass configurations are included. While our sampling ensures coverage up to at least $2\,M_\odot$  by design—through the choice of three uniformly distributed mass bands—this does not ensure that the model is exposed to the true maximum mass. As a result, any quantitative inference of $n_{c,max}$ would be inherently limited. A more targeted study of this aspect is left for future work.
While in this work we focused primarily on a qualitative proof of concept, future work will aim to develop a more systematic quantitative validation framework, exploring additional metrics beyond relative residuals and coverage tests.
\section*{ACKNOWLEDGMENTS} 

V.C. expresses sincere gratitude to the FCT for their generous support through Ph.D. grant number 2024.00311.BD. This work was partially supported by national funds from FCT (Fundação para a Ciência e a Tecnologia, I.P, Portugal) under the projects 2022.06460.PTDC with the  DOI identifier 10.54499/2022.06460.PTDC, and UIDB/04564/2020 and UIDP/04564/2020, with DOI identifiers 10.54499/UIDB/04564/2020 and 10.54499/UIDP/04564/2020, respectively, by the European Union-Next Generation EU, Mission 4 Component 1 CUP J53D23001550006 with the PRIN Project No. 202275HT58, and the Polish National Science Center OPUS grant no. 2021/43/B/ST9/01714.

\onecolumngrid
\appendix

\section{Implementation Details}
\label{ssec:implementation}

Our architecture uses the following components:
\begin{itemize}
\item \textbf{Software}: Implemented in \texttt{PyTorch} \cite{NEURIPS2019_9015} with the \texttt{nflows} library \cite{nflows}.
\item \textbf{Flow Design}:
\begin{itemize}
\item Number of flow transformations: 16,
\item 3 ResNet blocks \cite{he2016deep} per flow with 120 hidden units,
\item Exponential Linear Unit (ELU) activation  functions \cite{clevert2015fast} for smooth gradients.
\end{itemize}
\item \textbf{Optimization}:
\begin{itemize}
\item Adam optimizer \cite{kingma2014adam} with learning rate $10^{-4}$,
\item Batch size of 128, trained for 1000 epochs,
\item We set $\lambda = 0.7$ based on empirical evaluation. Among the tested values (0.4, 0.7, and 0.9), $\lambda = 0.7$ consistently yielded the best overall performance in terms of reconstruction accuracy and calibration.
\end{itemize}
\end{itemize}
\section{Calibration of Uncertainty Estimates}\label{appendix:coverage}

To complement the main analysis, we assess the calibration of the model’s predicted uncertainties by evaluating the empirical coverage probabilities across different datasets and physical quantities. These results support the statistical reliability of the CI used throughout the main text.
\subsection{Coverage for the PT Dataset}\label{sec:cov_pt}

\begin{figure}[H]
    \centering
    \includegraphics[width=0.4\linewidth]{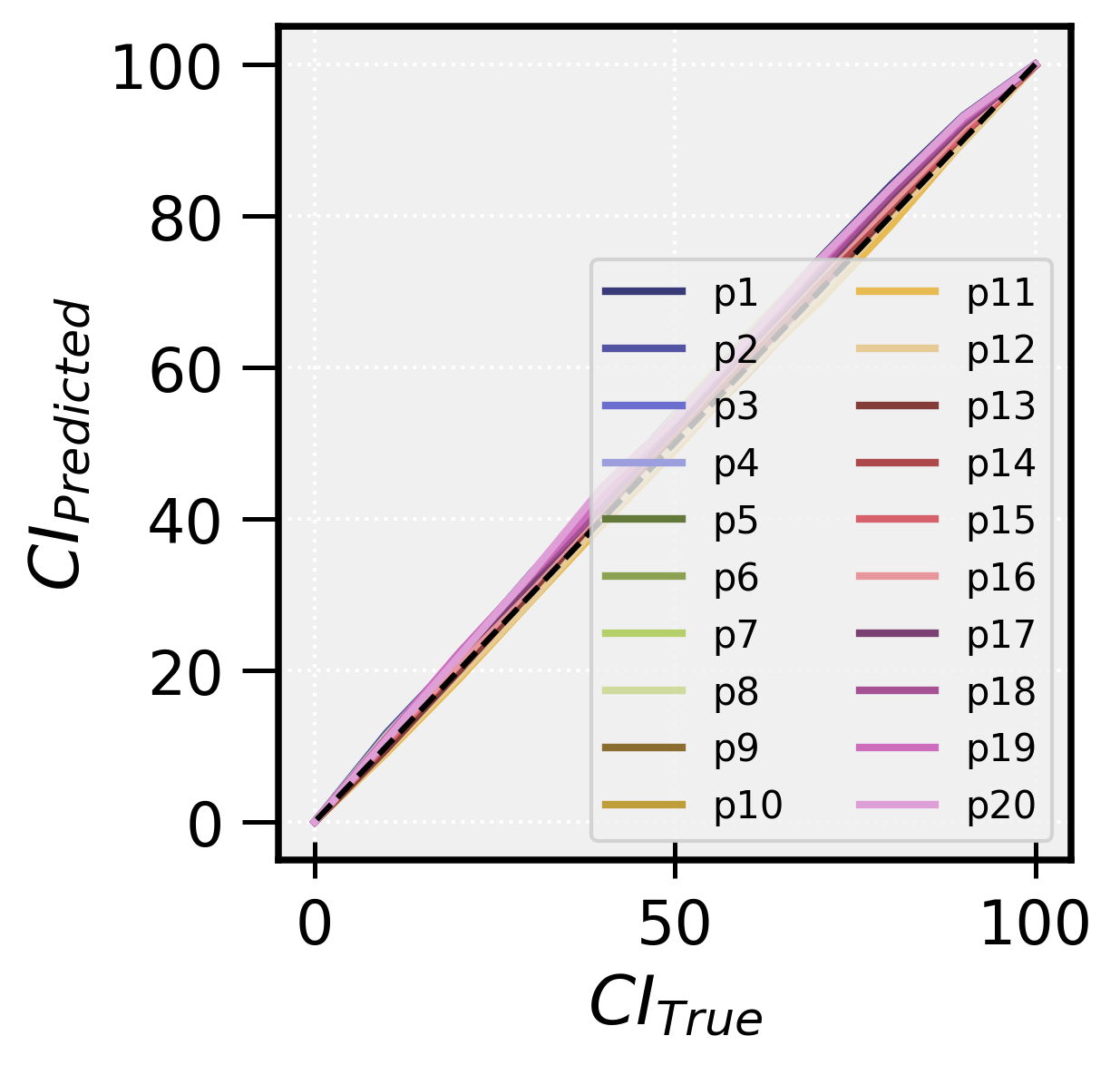}
    \includegraphics[width=0.4\linewidth]{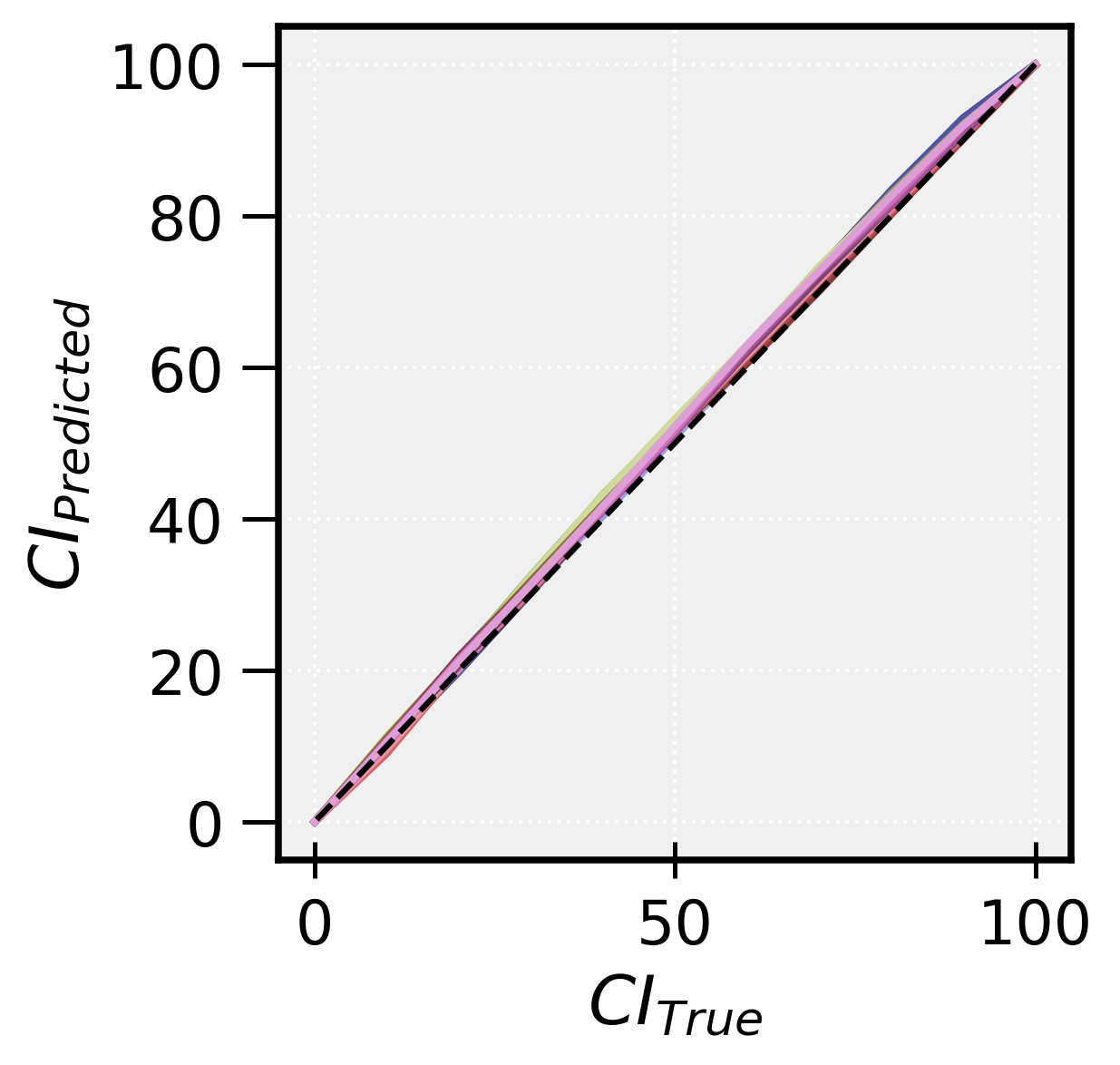}
    \caption{
    Empirical coverage probability for pressure values in the PT dataset. Left panel: \( R_2 \) dataset. Right panel: \( R\Lambda_2 \) dataset. Each curve shows the proportion of true pressure values within the predicted credible intervals, as a function of the nominal confidence level.
    }
    \label{fig:rel_poly}
\end{figure}

Figure~\ref{fig:rel_poly} shows the empirical coverage probability curves for the 20 pressure components across two representative datasets in the PT set: \( R_2 \) (left) and \( R\Lambda_2 \) (right). Each curve represents the fraction of true pressure values that fall within the model's predicted credible intervals, plotted as a function of the nominal confidence level.

Both models exhibit excellent calibration, with empirical coverage closely following the diagonal (ideal) line, indicating that the predicted uncertainties are statistically reliable across the full range of pressure values.

\subsection{Coverage for the GP Dataset}\label{sec:cov_gp}

\begin{figure}[H]
    \centering
    \includegraphics[width=0.3\linewidth]{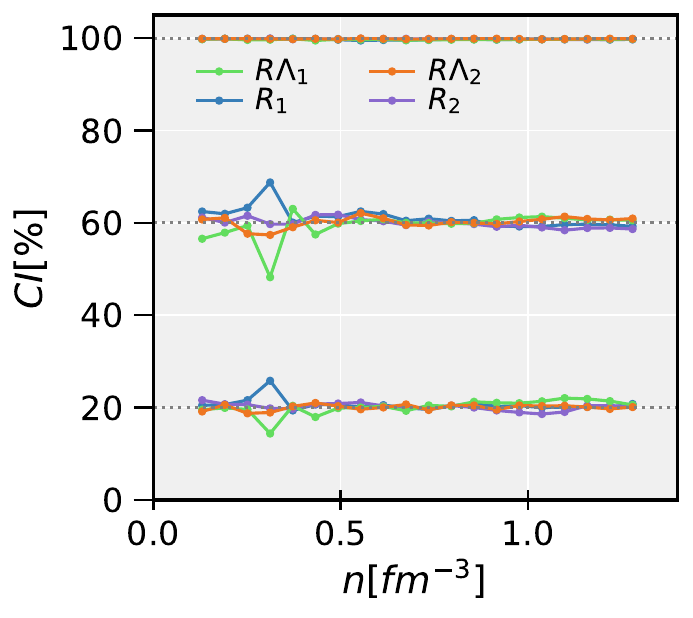}
    \includegraphics[width=0.3\linewidth]{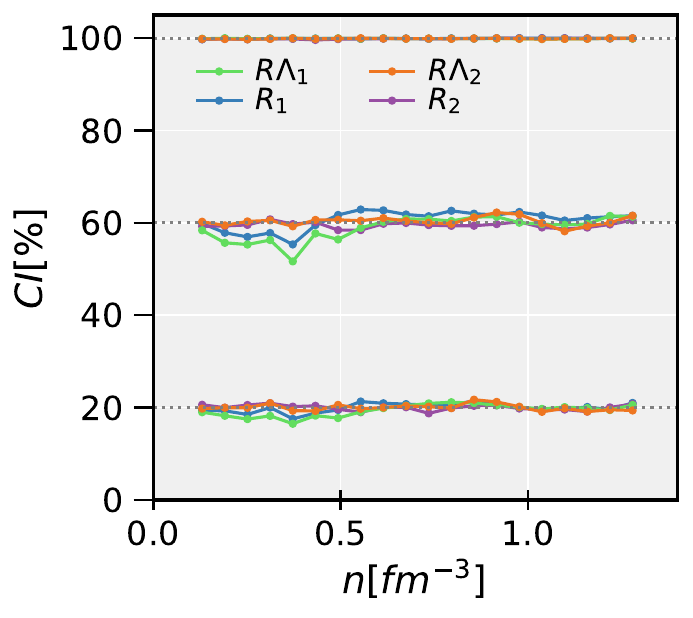}
         \includegraphics[width=0.3\linewidth]{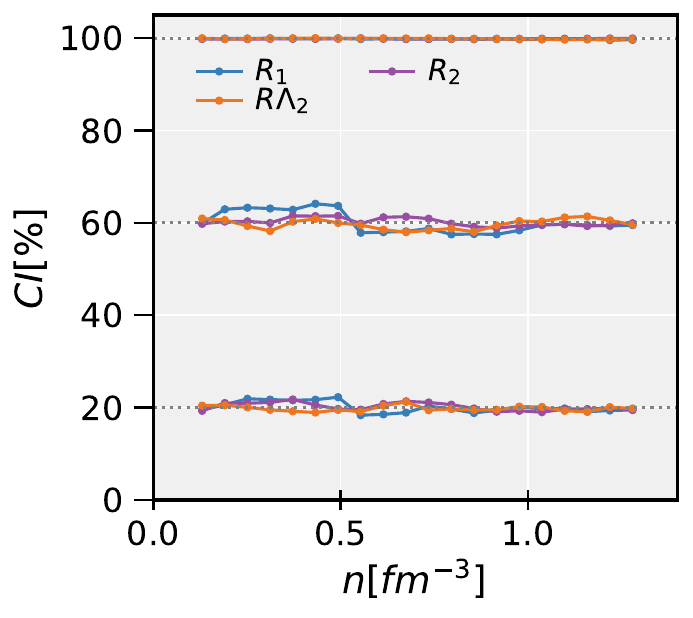}
    \caption{
    Empirical coverage probability at 20\%, 60\%, and 100\% confidence levels for the GP dataset. {Left panel:} Pressure \( p(n) \); {Middle panel:} Squared speed of sound \( c_s^2(n) \); {Right panel:} Trace anomaly \( \Delta(n) \). The x-axis represents the baryon density, and the y-axis shows the fraction of samples whose true values fall within the predicted credible intervals. Colors correspond to different datasets.
    }
    \label{fig:rel_gp}
\end{figure}

Fig.~\ref{fig:rel_gp} displays the empirical coverage curves for the GP dataset, evaluated at three confidence levels (20\%, 60\%, and 100\%) for the pressure \( p(n) \) (left plot), squared speed of sound \( c_s^2(n) \) (center plot), and trace anomaly \( \Delta (n) \)  (right plot), across 20 density points.

For pressure, the model shows well-calibrated uncertainty estimates across most of the density range. A small oscillation appears near \( n \approx 0.4 \, \text{fm}^{-3} \), possibly caused by rapid local changes in the EoS slope. Interestingly, this region coincides with the first local maximum of the speed of sound, suggesting that the model's uncertainty estimation is more sensitive to higher-order structural features in the EoS.

The center and right plots show coverage probabilities for \( c_s^2(n) \) and   \( \Delta (n) \). While coverage is slightly underestimated at low densities—likely due to localized fluctuations—the model maintains good calibration across the rest of the range. This reflects robust uncertainty estimation, particularly in regions where the EoS is smoother and better constrained.

\section{Correlation with Maximum Central Density \( n_{c,\mathrm{max}} \)}\label{sec:AP.corre}

We investigate whether the point of steepest growth in model uncertainty—quantified via the slope of the dispersion curve—correlates with the true maximum central density \( n_{c,\mathrm{max}} \) of each EoS. Due to the wide dynamic range of pressure values, we use a normalized dispersion for pressure, whereas for quantities like the squared speed of sound and the trace anomaly, the raw 90\% CI width is used.

\vspace{1em}
We define the dispersion functions as
\begin{align}
\delta_p(n) &= \frac{p_{90\%\mathrm{CI}}(n)}{\overline{p}(n)} \;,\\
\delta_{c_s^2}(n) &= c_{s,90\%\mathrm{CI}}^2(n)\;, \\
\delta_{\Delta}(n) &= \Delta_{90\%\mathrm{CI}}(n)\;.
\end{align}

The estimated transition density corresponding to the point of maximum uncertainty growth is given by:
\begin{equation}
n_{\mathrm{est}}^{(x)} = \underset{n}{\arg\max} \left( \frac{d}{dn} \delta_x(n) \right),
\quad x \in \{p, c_s^2, \Delta\} \;.
\end{equation}

\begin{figure}[!hbt]
    \centering
    \includegraphics[width=0.27\linewidth]{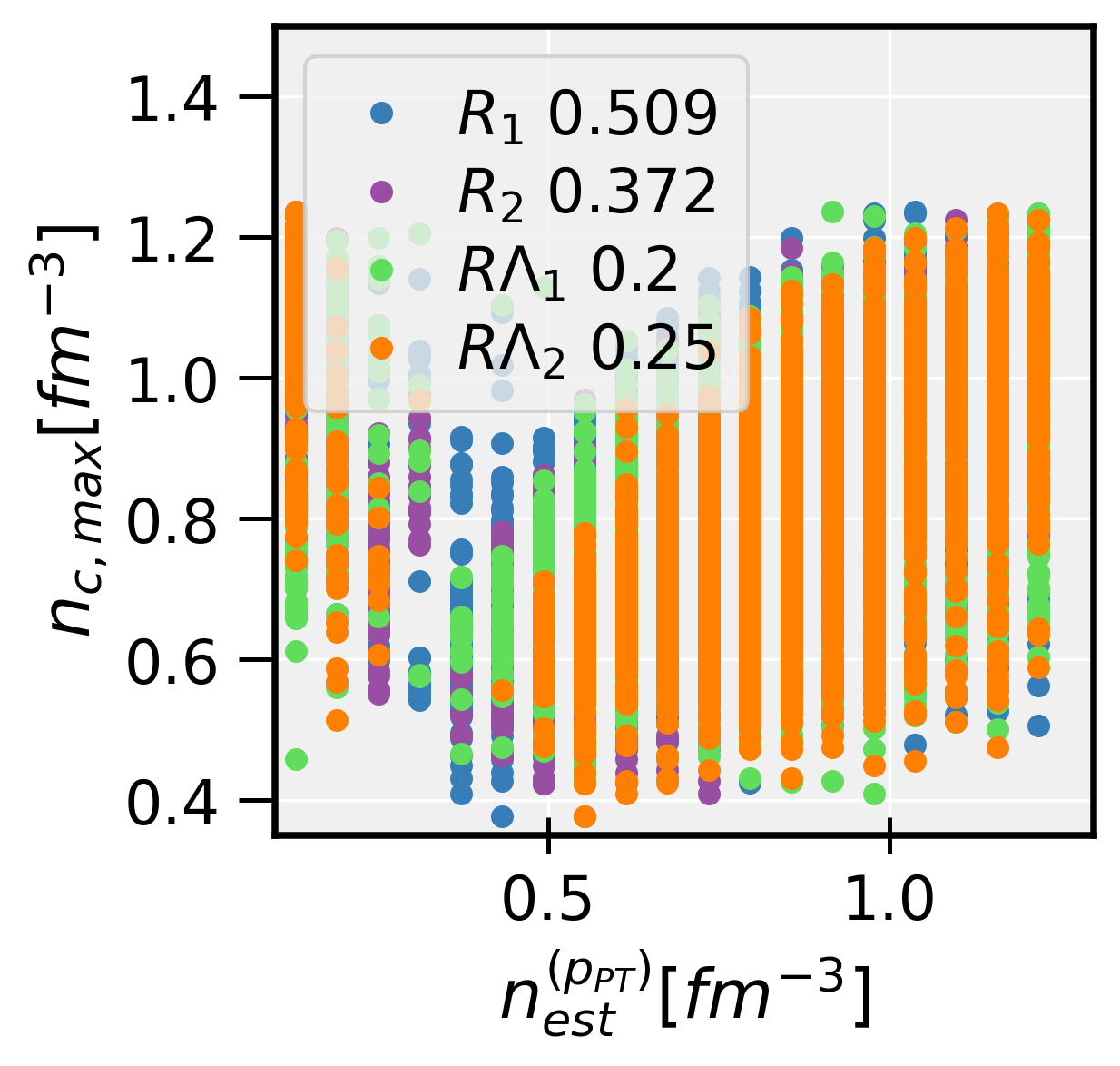}
    \includegraphics[width=0.23\linewidth]{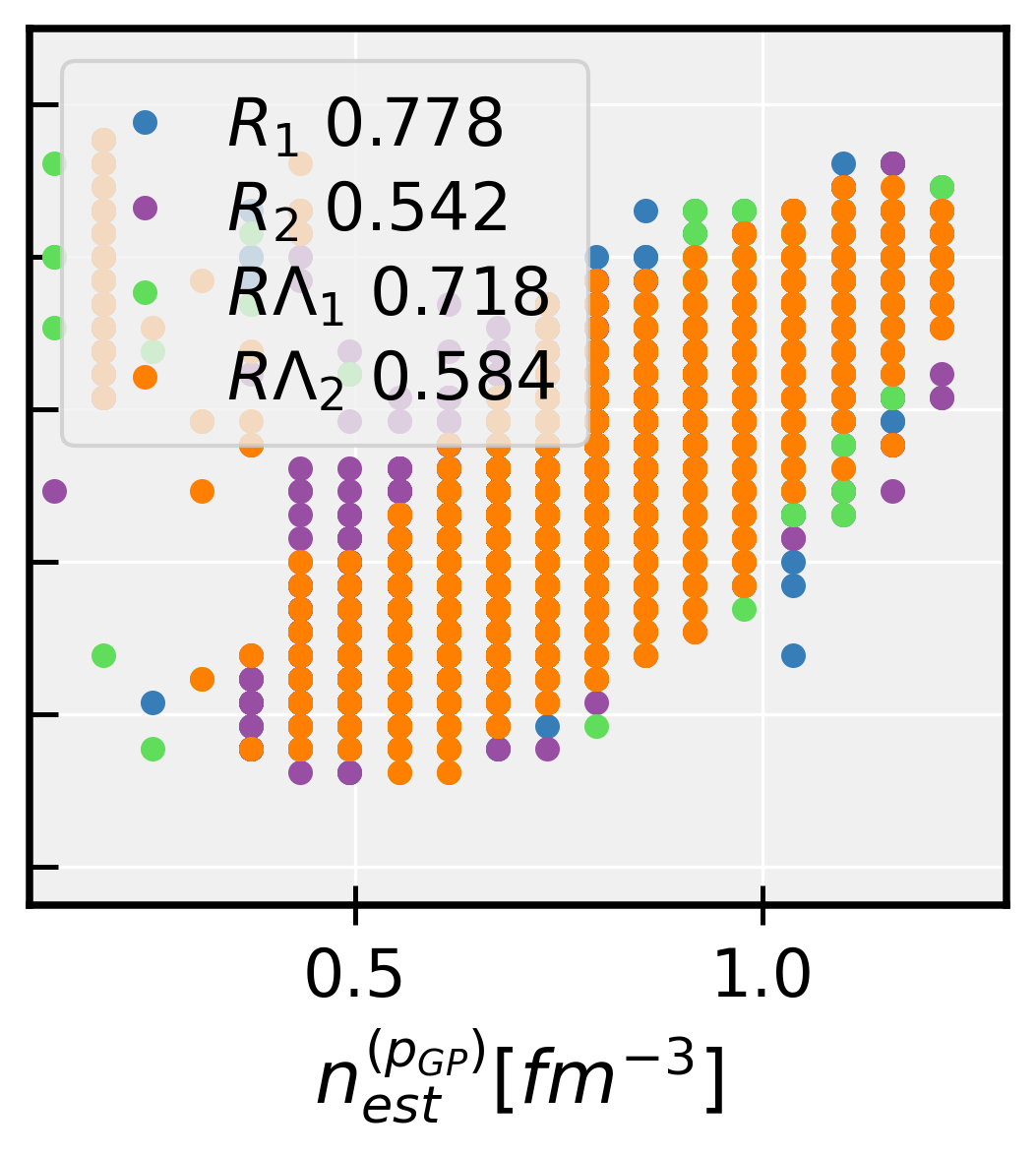}
    \includegraphics[width=0.23\linewidth]{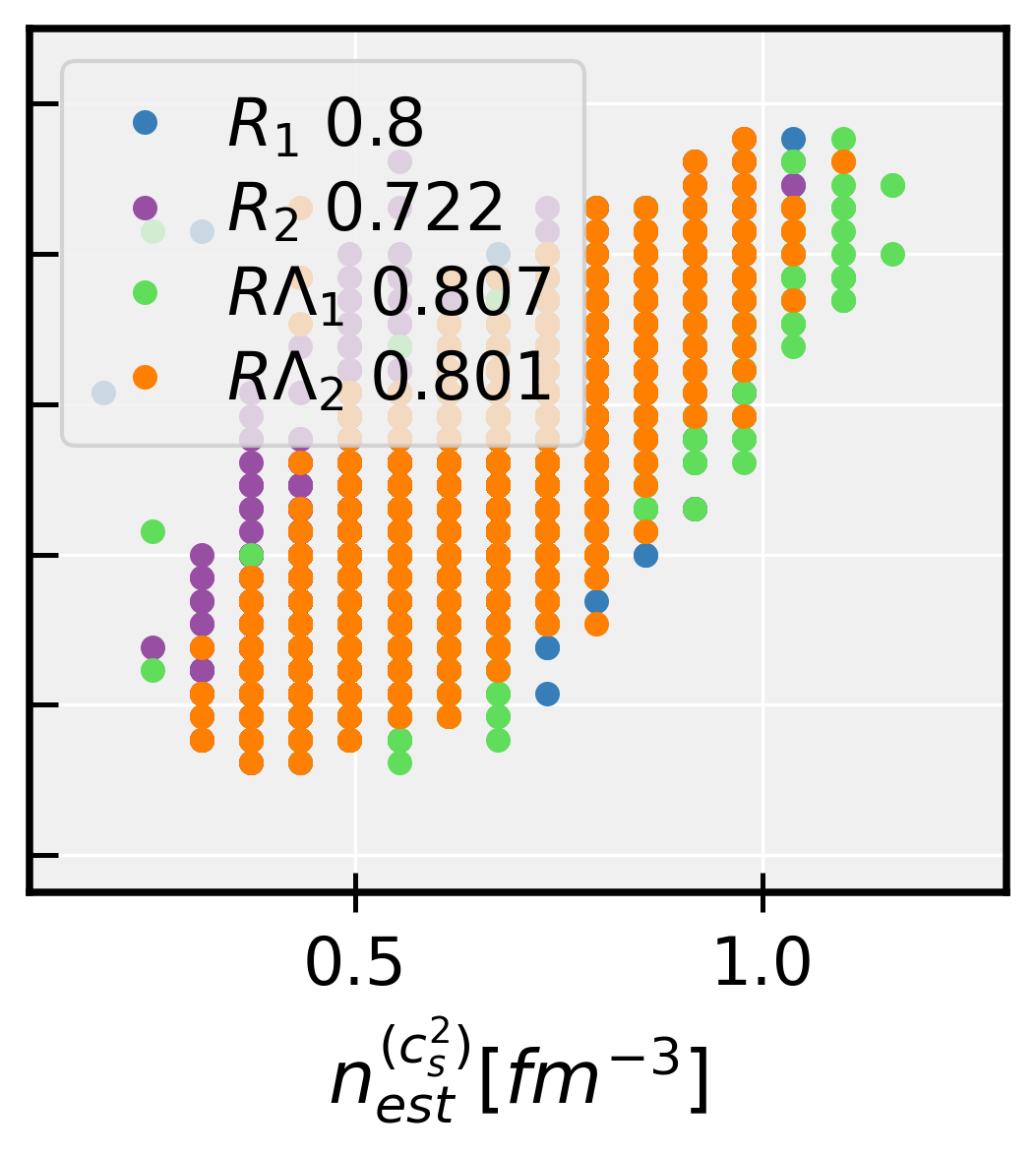}
    \includegraphics[width=0.23\linewidth]{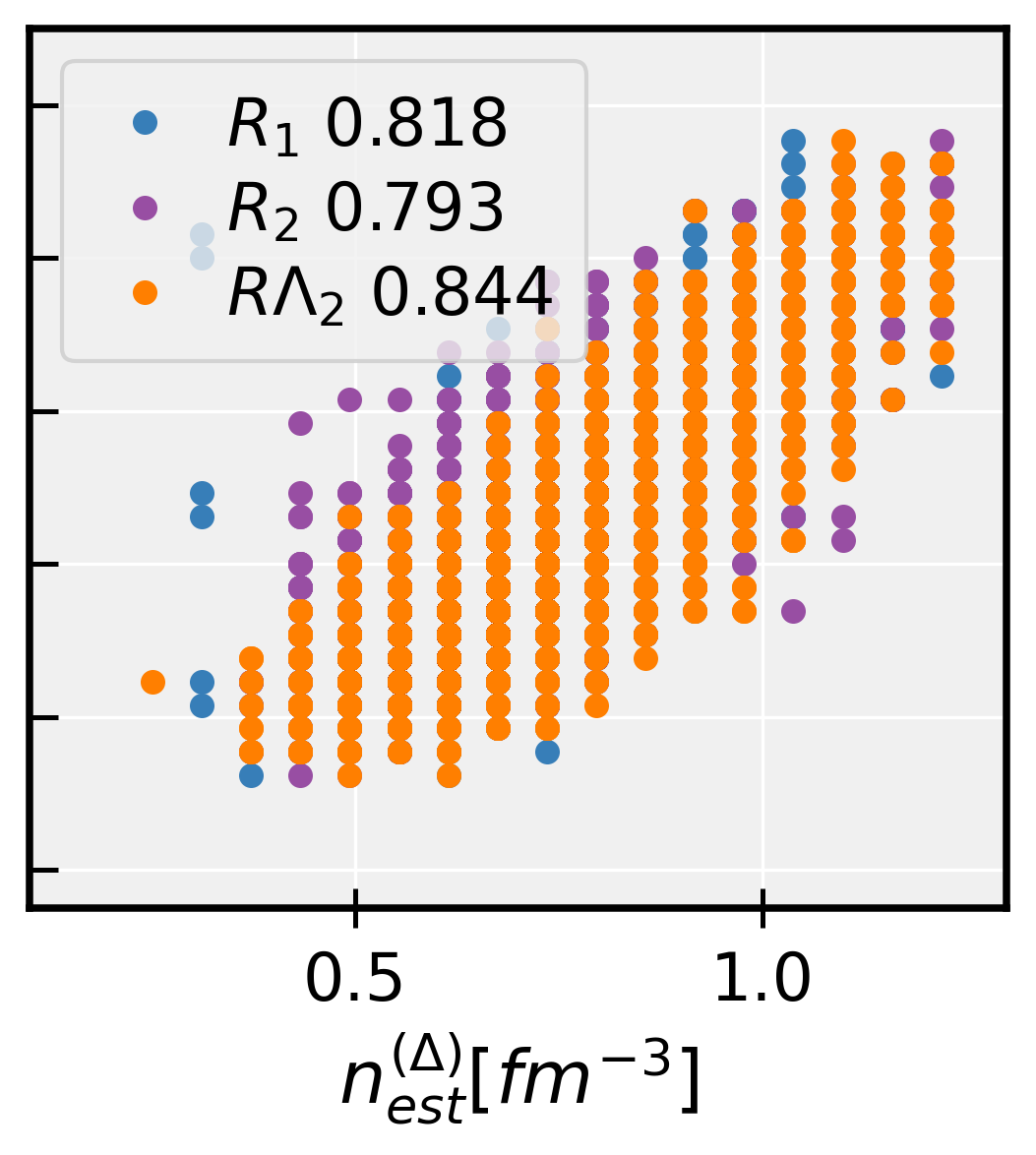}
        \caption{Correlation between the estimated transition density \( n_{\mathrm{est}}^{(x)} \) (density of maximum dispersion slope) and true maximum central density \( n_{c,\mathrm{max}} \) across the four dataset configurations in Sec.~\ref{dataset}. Pearson correlation coefficients are shown in the legend of each plot. Each plot corresponds to a different study quantity: leftmost, pressure (PT), center-left pressure (GP), center-right, squared speed of sound  \( c_s^2 \) (GP), and rightmost, trace anomaly \( \Delta \) (GP).}
    \label{fig:correla}
\end{figure}

This procedure is applied separately to each EoS in the test set across the four dataset configurations and three diagnostic quantities. The resulting values of \( n_{\mathrm{est}}^{(x)} \) are then compared with the true \( n_{c,\mathrm{max}} \), as shown in Fig.~\ref{fig:correla}.

The correlation between the estimated transition density and true maximum central density is consistently strong for the GP-based datasets—especially when using the squared speed of sound or trace anomaly—with Pearson coefficients exceeding 0.7. In contrast, the PT dataset exhibits weaker correlations, likely due to the intrinsic piecewise structure and derivative discontinuities of PT EoSs. The presence of discontinuities or sharp transitions in the EoS can create artificial changes in the dispersion slope, which leads to the dispersion predicted by the model being less sensitive to the maximal central density.

\twocolumngrid
\bibliographystyle{apsrev4-1}
\bibliography{biblio}

\end{document}